\documentclass[traditabstract,twocolumn]{aa}
\usepackage[varg]{txfonts}

\usepackage{bm}

\usepackage{graphicx}
\usepackage{natbib}
\usepackage{color}
\usepackage{subfigure}
\usepackage{xspace}
\usepackage{amssymb}
\usepackage{xspace}
\usepackage{url}
\usepackage{wasysym}
\newcommand\simlt{\lower.5ex\hbox{$\; \buildrel < \over \sim \;$}}

\def\grays{$\gamma$-rays\xspace}
\def\fermi{{\it Fermi}-LAT\xspace}
\def\planck{{\it Planck}\xspace}

\def\gray{$\gamma$-ray\xspace}

\begin{document}

\title{Giant lobes of Centaurus A  as seen in radio and gamma-ray images obtained with  the 
\fermi and \planck satellites}
\author{Xiao-na Sun\inst{1}
\and Rui-zhi Yang\inst{1}
\and Benjamin Mckinley\inst{2,3}
\and Felix Aharonian\inst{4,1,5}}
\institute{Max-Planck-Institut f{\"u}r Kernphysik, P.O. Box 103980, 69029 Heidelberg, Germany.
\and School of Physics, University of Melbourne, Parkville, VIC 3010, Australia
\and ARC Centre of Excellence for All-sky Astrophysics (CAASTRO)
\and Dublin Institute for Advanced Studies, 31 Fitzwilliam Place, Dublin 2, Ireland.
\and MEPHI, Kashirskoe shosse 31, 115409 Moscow, Russia
}%



\abstract {
The  \gray  data of \fermi  on the giant lobes of  Centaurus A  
are  analysed together with the  high frequency  radio data 
obtained  with  the \planck  satellite.  
The large $\gamma$-ray photon statistics, accumulated during seven years of observations, and the  recently updated \fermi collaboration software tools allow substantial extension of the detected \gray emission towards higher energy, up to 30 GeV, and lower energy, down to 60~MeV.  
Moreover, the new \gray data allow us to explore the spatial features of \gray emission of the lobes. 
For the north lobe, we confirm, with higher statistical significance, our earlier finding on the extension of \gray emission beyond the radio image.
Moreover, the new analysis reveals significant spatial variation of \gray spectra from both lobes. On the other hand, the \planck observations at microwave frequencies contain important information on spectra of synchrotron emission in the cutoff region, and thus allow model-independent derivation of the strength of the magnetic field and the  distribution of relativistic  electrons based  on the combined \gray and radio data. The interpretation of  multiwavelength spectral energy  distributions (SEDs) of the lobes within a pure  leptonic model requires strong enhancement of the magnetic field at the edge of the south lobe. Alternatively, a more complex, leptonic-hadronic model of the gamma-ray emission, postulating a non-negligible contribution of the $\pi^0$-decay component at highest energies, can explain the \gray data with a rather homogeneous distribution of the magnetic field over the giant lobes. 
 }

\titlerunning{giant lobes of Centaurus A}
\maketitle

\section{Introduction}
\label{sec:intro}
Centaurus A (Cen A) is a Fanaroff-Riley type I (FR I) radio galaxy that is hosted by the  massive elliptical galaxy NGC 5128 \citep[see][]{1998A&ARv...8..237I}. It is the closest  radio galaxy located at a distance of $3.8 \pm 0.1~{\rm Mpc}$ (\citep{2010PASA...27..457H}; $1\arcmin \simeq 1.14 $kpc). Cen A contains a  central black hole of mass  
$ (5.5 \pm 3.0)  \times 10^{7}$\,$M_\odot$ \citep{2009MNRAS.394..660C}. 
The dynamical age  of the galaxy is $\sim$ 0.5 $-$1.5 Gyr \citep{2013A&A...558A..19W, 2014MNRAS.442.2867W, 2014NJPh...16d5001E}. 
At radio frequencies, a pair of giant lobes are visible, extending from the core to the north and south with an  angular size of $\sim 10^ \circ$ ($\sim 600$ kpc in projection) \citep{1958AuJPh..11..517S, 1983ApJ...273..128B}; Because of its unique proximity and complex morphology, the giant lobes have been extensively studied in both radio \citep[e.g.][]{1997A&AS..121...11C, 2013MNRAS.432.1285S,2000A&A...355..863A, 2009MNRAS.393.1041H, 2013MNRAS.436.1286M} and \gray bands \citep[e.g.][]{2010Sci...328..725A, 2012A&A...542A..19Y}.

High energy \gray emission in the outer lobes is believed to be produced through the inverse Compton (IC) channel, when the relativistic electrons yield an upscatter of low energy background photons, including the cosmic microwave background (CMB) and ubiquitous extragalactic background light (EBL), to MeV-GeV energies \citep{2010Sci...328..725A, 2012A&A...542A..19Y}. This provides us a unique opportunity to map the spatial and energy distribution of relativistic electrons in this source. Furthermore, by comparing the radio/microwave and \gray emissions, we can obtain unambiguous information on the magnetic fields.   

Owing to the accumulative photon statistics and the recently improved software tools of \fermi, we can extend the original analysis of \citet{2012A&A...542A..19Y} to lower and higher energies and investigate the spatial variation of \gray spectra. In such a low magnetic field the electrons that produce GeV \grays via IC scattering, have much higher energies than those which produce radio/microwave radiation via synchrotron radiation. The \planck satellite provides high sensitivity data with full sky coverage extending from 30 GHz to 853 GHz. These frequencies minimise the energy gap between the two electron populations responsible for radio and \grays emissions.  
 
In this paper, we present a detailed analysis of the broadband emission of the lobes of  Cen A using \gray data from \fermi and microwave data from \planck. In Section~\ref{sec:fermi_analy}, we perform  the analysis  of \fermi data. In Section~\ref{sec:planck_analy}, we analyse the MHz-range data from radio telescopes and microwave data from \planck . 
In Section~\ref{sec:fitting}, we fit the broadband SEDs of the lobes  within 
the pure leptonic and more complex leptonic-hadronic models. 
We discuss the results in Section~\ref{sec:conc}.

\section{\fermi data analysis}\label{sec:fermi_analy}

\begin{figure*}
\centering
\includegraphics[width=0.5\linewidth]{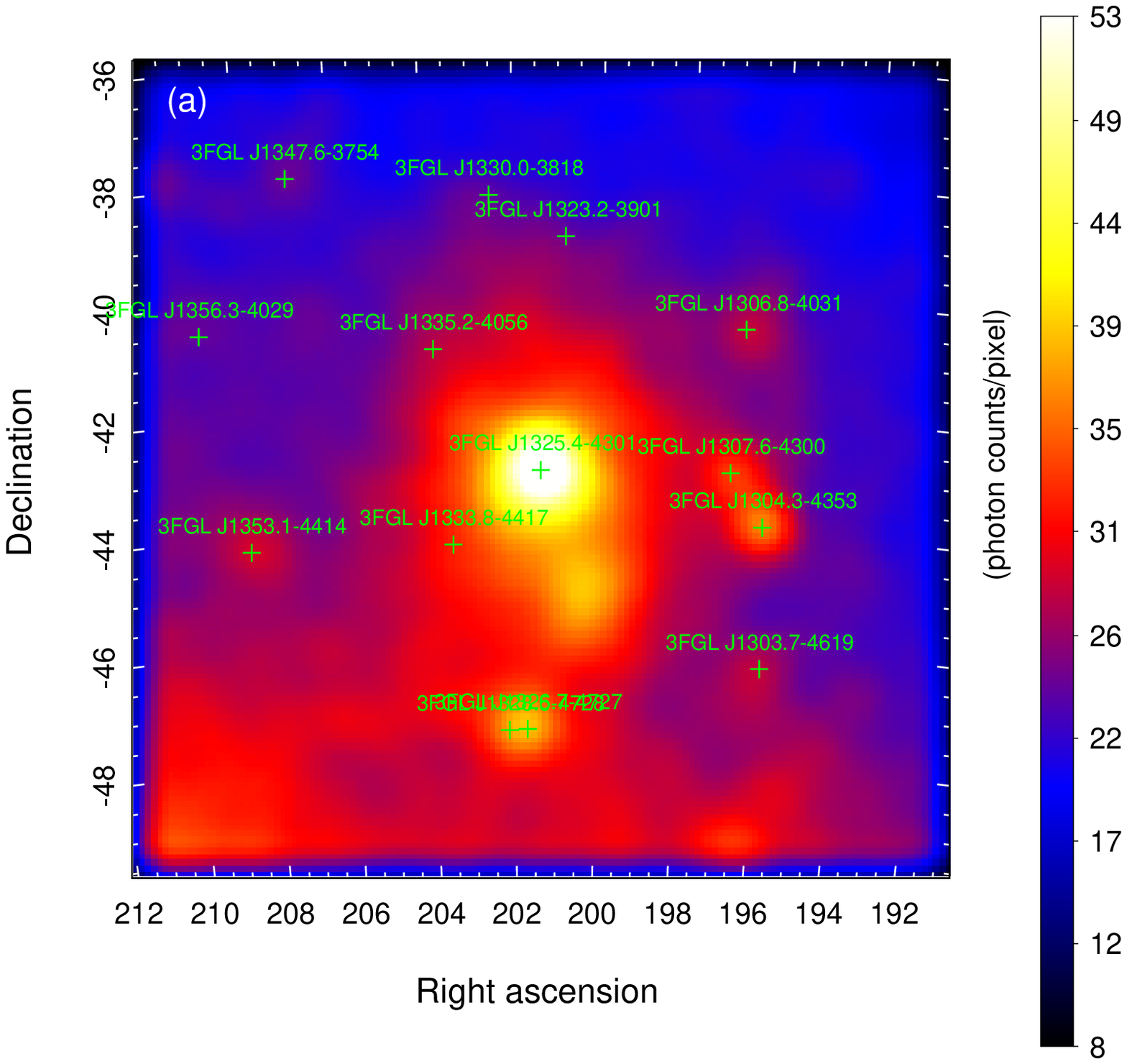}\includegraphics[width=0.5\linewidth]{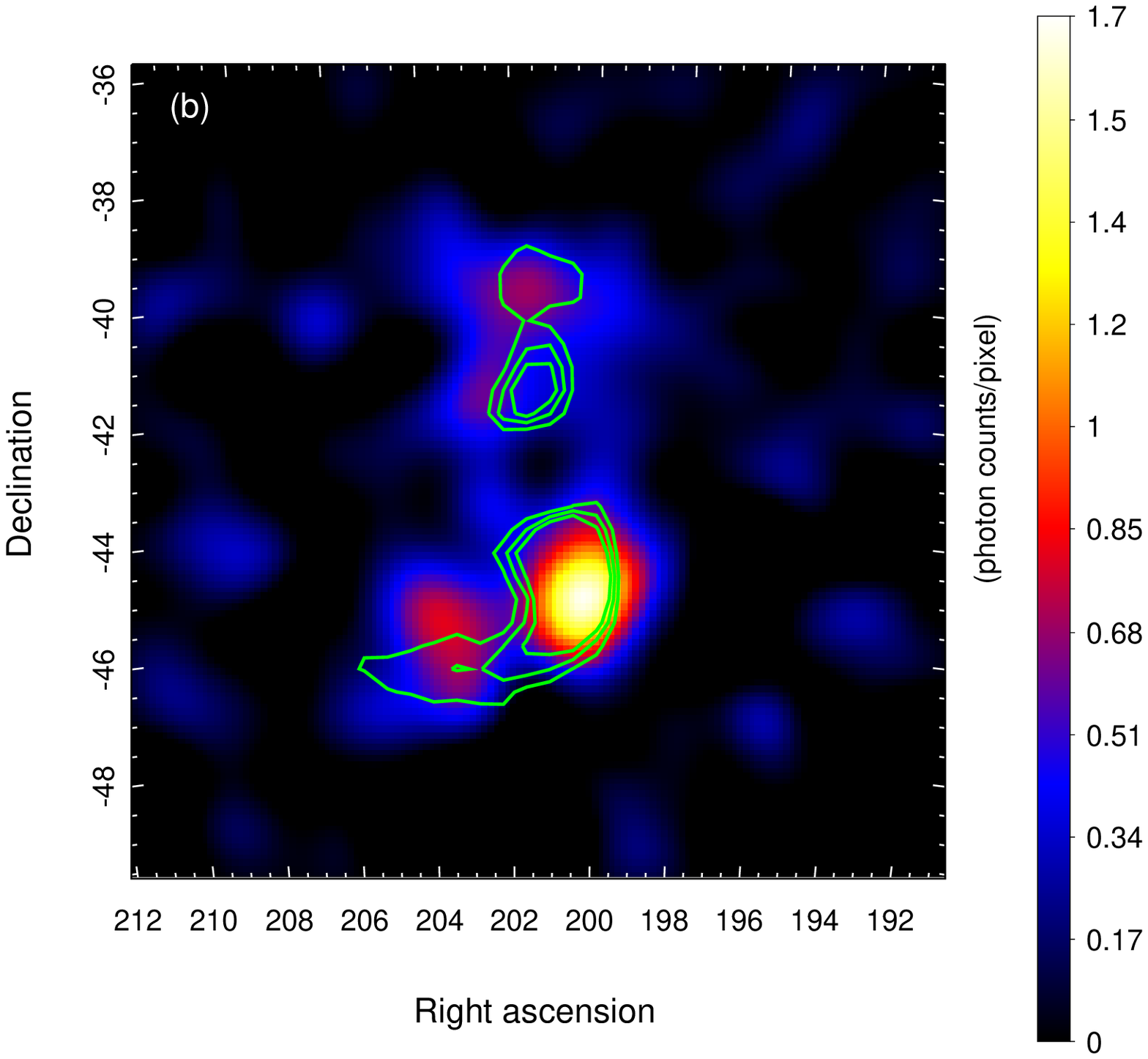}
\caption{(a) LAT counts map of the $14^ \circ \times 14^ \circ$ ROI. The counts map is smoothed with a Gaussian 
of kernel $0.7^ \circ$. The green crosses indicate the position of the point-like sources within $7^ \circ$ of Cen A.
(b) Residual maps of lobes after subtracting the diffuse background, the point-like sources, and the Cen A core. The green contours indicate Planck 30 GHz lobe contours.}
\label{fig:1}
\end{figure*}

The analysis of this section includes the  \fermi data  from the directions of the two giant radio lobes. 
We selected observations from August 4, 2008 (MET 239557417) until June 27, 2015 (MET 457063584) and used 
photons in the energy range between $60~ {\rm MeV}$ and $30~ {\rm GeV}$.  A $14^ \circ \times 14^ \circ$ square region centred at the location of Cen A (RA = $201^{\circ}21'54''$, Dec = $-43^{\circ}1'9''$) was selected as the region of interest (ROI). We selected both the front and back converted photons.
To reduce the background contamination from  the Earth's albedo, the events from directions $> 90^ \circ$ were excluded from the analysis.
We adopted the version P8R2\_SOURCE\_V6  of the instrument response function (IRF) provided by the 
\fermi collaboration. The binned likelihood analysis implemented in science tool {\it gtlike} was used to evaluate the spectrum. 

To define the initial source list, we used the four-year catalogue (3FGL) \citep{3fgl} by running the make3FGLxml script\footnote{\url{http://fermi.gsfc.nasa.gov/ssc/data/analysis/user/make3FGLxml.py}}. 
In the initial list, the spectral parameters of point-like sources within the ROI were left as free parameters. Also, we used the default spatial template for giant lobes provided by Fermi collaboration. 
We used the \emph{Fermi} models of Galactic diffuse and isotropic emission provided by the \emph{Fermi} team\footnote{\url{http://fermi.gsfc.nasa.gov/ssc/data/access/lat/BackgroundModels.html}} for the foreground components. 
In the fitting, the normalisations  of both components were left as free parameters.

\subsection{Spatial analysis}\label{sec:spatial_analy}
For morphology studies, we also built our own templates directly from the \gray residual maps. We applied  {\it gtlike} with the initial source list and derived a fitted model. Then we produced the residual maps by removing the contribution from the diffuse background and all catalogue sources except the giant lobes.   
We also masked  the inner $1^ \circ$ nucleus of Cen A to prevent the contamination from that region. Finally, we divided the residual maps into north and south lobes. Generally speaking, high energy maps with higher angular resolution are more suitable for the spatial analysis, but low statistics in the higher energy range may prevent any improvement. To address this problem, we applied the  procedure described above to the energy range  $> 1000~ {\rm MeV}$ and $> 100~ {\rm MeV}$, respectively. The resulting spatial templates are labelled as T1 ($> 1000~ {\rm MeV}$) and T2 ($> 100~ {\rm MeV}$). 

We used the residual templates T1 and T2, as well as the default spatial templates provided by the \emph{Fermi} team\footnote{\url{http://fermi.gsfc.nasa.gov/ssc/data/analysis/scitools/extended/extended.html}} (T3), to model 
the giant lobes, and applied  {\it gtlike} to  three models in the energy range above $100~ {\rm MeV}$. 
The resulting TS and -log(Likelihood) values are listed in Table~1. In the case of template T1, the core of Cen A is clearly visible with a test statistic of TS $> 7000$, corresponding to a detection significance  of $84\sigma$. Extended emission to the north and south of the lobes of Cen A is detected with significances of TS $> 450$ $(21\sigma)$ and TS $> 1500 $ $(39\sigma)$, respectively. The -log(Likelihood) for T1  is significantly smaller than that of T2 and T3. Therefore, for our analysis, we selected  template T1. 
The \fermi counts map  produced  for the $> 1000~ {\rm MeV}$ data set, is shown in Figure~\ref{fig:1}(a);
the green crosses show the position of point-like sources from the 3FGL catalogue within the ROI. The corresponding residual image (template T1) is shown in Figure~\ref{fig:1}(b). For comparison, the Planck 30 GHz lobe contours (green contours) are  also plotted in the image.  It can be seen  that both the north and south lobes of the HE \gray emission extend beyond the radio lobes. 

\subsection{Spectral analysis}
\begin{figure}
\centering
\includegraphics[scale=0.5]{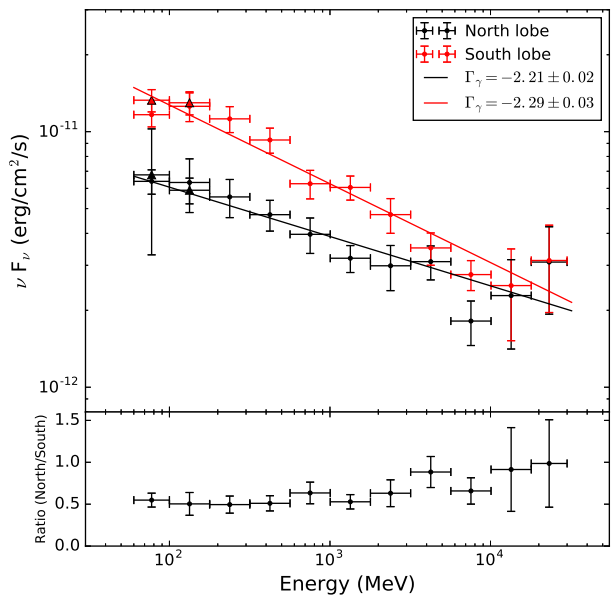}
\caption{Derived SEDs of the north lobe and south  lobe, with power-law fits. The corresponding photon indices are $\Gamma_{\rm N}$ and $\Gamma_{\rm S}$. The ratio of the north and south  fluxes  is shown in the bottom panel. For the first two energy bins, the SEDs derived using a larger ROI are also shown as bold triangles.
}
\label{fig:2}
\end{figure}

To derive the SED, we divided the energy interval between $100~ {\rm MeV}$ and $30~ {\rm GeV}$ into ten equal bins in logarithmic space and used {\it gtlike} in each bin. To extend the spectral analysis to lower energies, we also included photons with energy between $60~ {\rm MeV}$ and $100~ {\rm MeV}$,  and regarded these as the first energy bin. We applied energy dispersion correction to this energy bin.  The significance of the signal detection in each energy bin 
exceeds $TS=8$ ($\sim 3\sigma$). The SEDs fitted with a power law are shown in Figure~\ref{fig:2}. Correspondingly, the photon indices of the north and south  lobes are $(2.21\pm0.02)$ and $(2.29\pm0.03)$, respectively. Within uncertainties, the index of the north  lobe is consistent with that in \citet {2012A&A...542A..19Y}, while  the index of the south lobe  is slightly smaller. The integral flux above $100~ {\rm MeV}$ is $(0.54 \pm 0.06) \times 10^{-7}~\rm ph~cm^{-2}~s^{-1}$ for the north  lobe and $(1.22 \pm 0.05) \times 10^{-7}~\rm ph~cm^{-2}~s^{-1}$ for the south  lobe. 

At low energy, the point spread function (PSF) of Fermi can be as large as $5^{\circ}$, in which case the $14^ \circ \times 14^ \circ$ ROI  used here may be not sufficient. We refit the first two energy bins using a $20^ \circ \times 20^ \circ$ ROI to see the influence of the limited ROI. These results are also shown in Figure~\ref{fig:2} and are consistent with those derived from the smaller ROI. 

\begin{figure}
\centering
\includegraphics[scale=0.5]{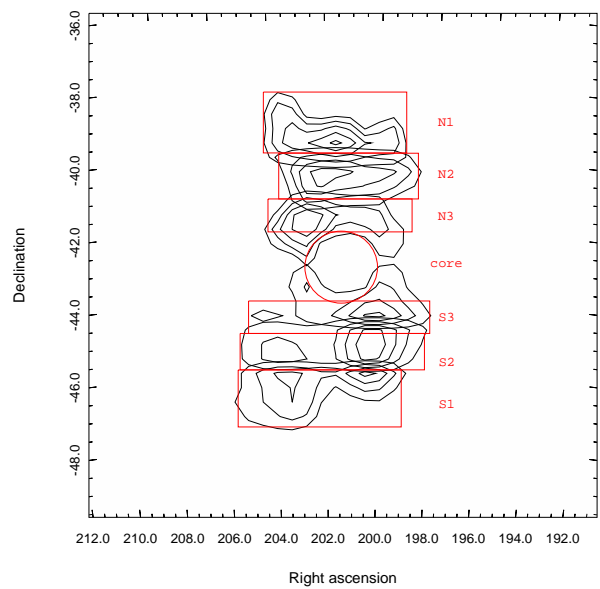}
\caption{Spectral extraction regions used for Figure~\ref{fig:4}.  The contours correspond to the  \gray image (>1000 MeV). The red  lines indicate  the 
regions of the radio and \planck aperture photometry. The contours inside the corresponding regions show the template
used for the extraction of the corresponding LAT spectrum.  N1, N2, and N3 are  the outer, middle, and inner regions of the north lobe, and S1, S2, and S3 are the outer, middle, and inner regions of the south lobe. The circle is the core region.}
\label{fig:3}
\end{figure}

To test the possible spectral variations from the outer regions of the lobes towards the central core, we split each lobe of template T1 into three parts. These regions are shown in Figure~\ref{fig:3}. The contours of the spectral extraction regions were used for the \fermi analysis and the red rectangles were used for the radio and Planck data aperture photometry. The outer, middle and inner regions of the north lobe are N1, N2 and N3, respectively, and S1, S2 and S3 are, the outer, middle and inner regions of the south lobe, respectively. The circle is the core region.

\begin{figure*}
\setlength{\abovecaptionskip}{1pt} 
\setlength{\belowcaptionskip}{1pt} 
\centering
\includegraphics[width=0.5\linewidth]{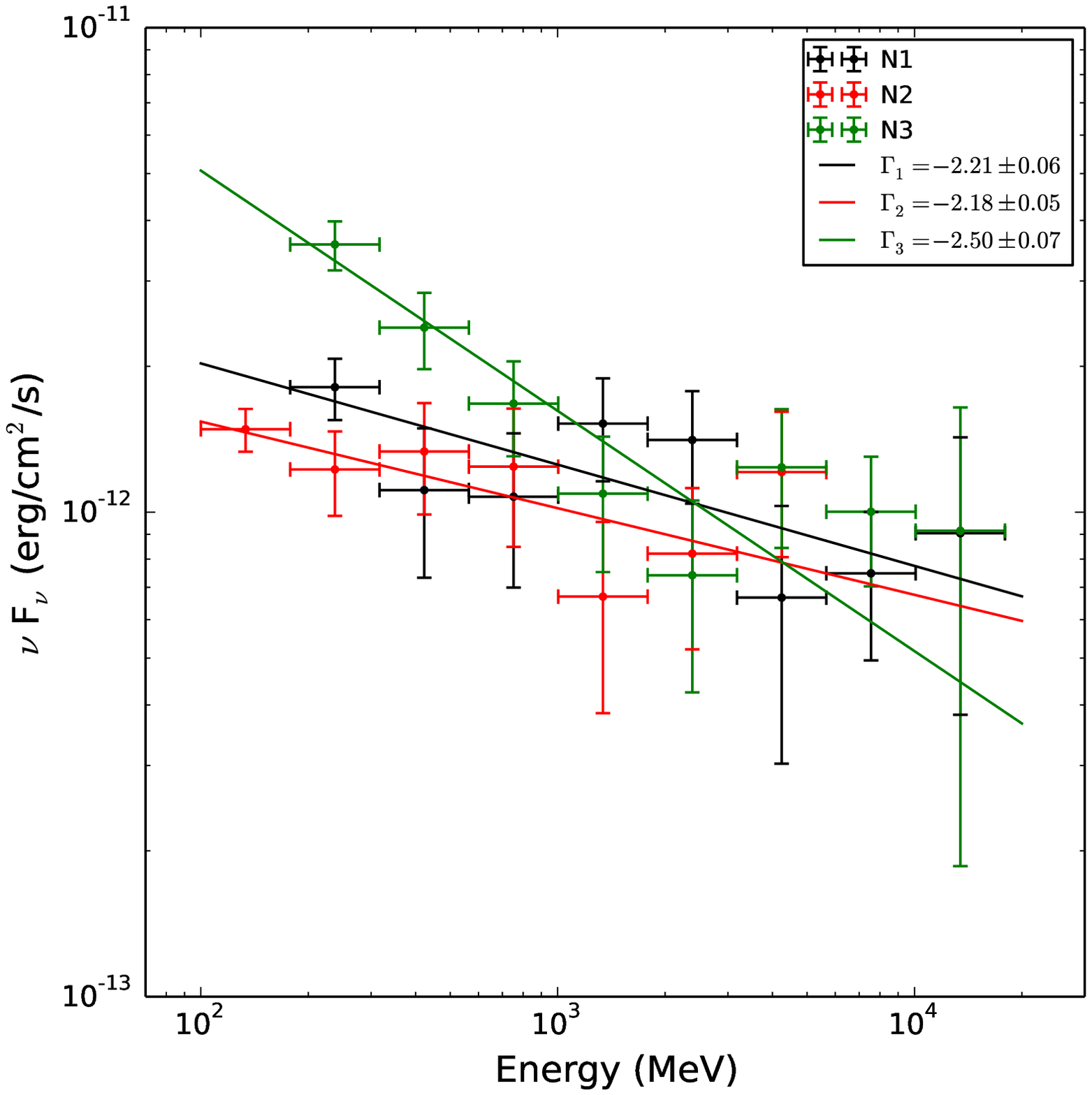}\includegraphics[width=0.5\linewidth]{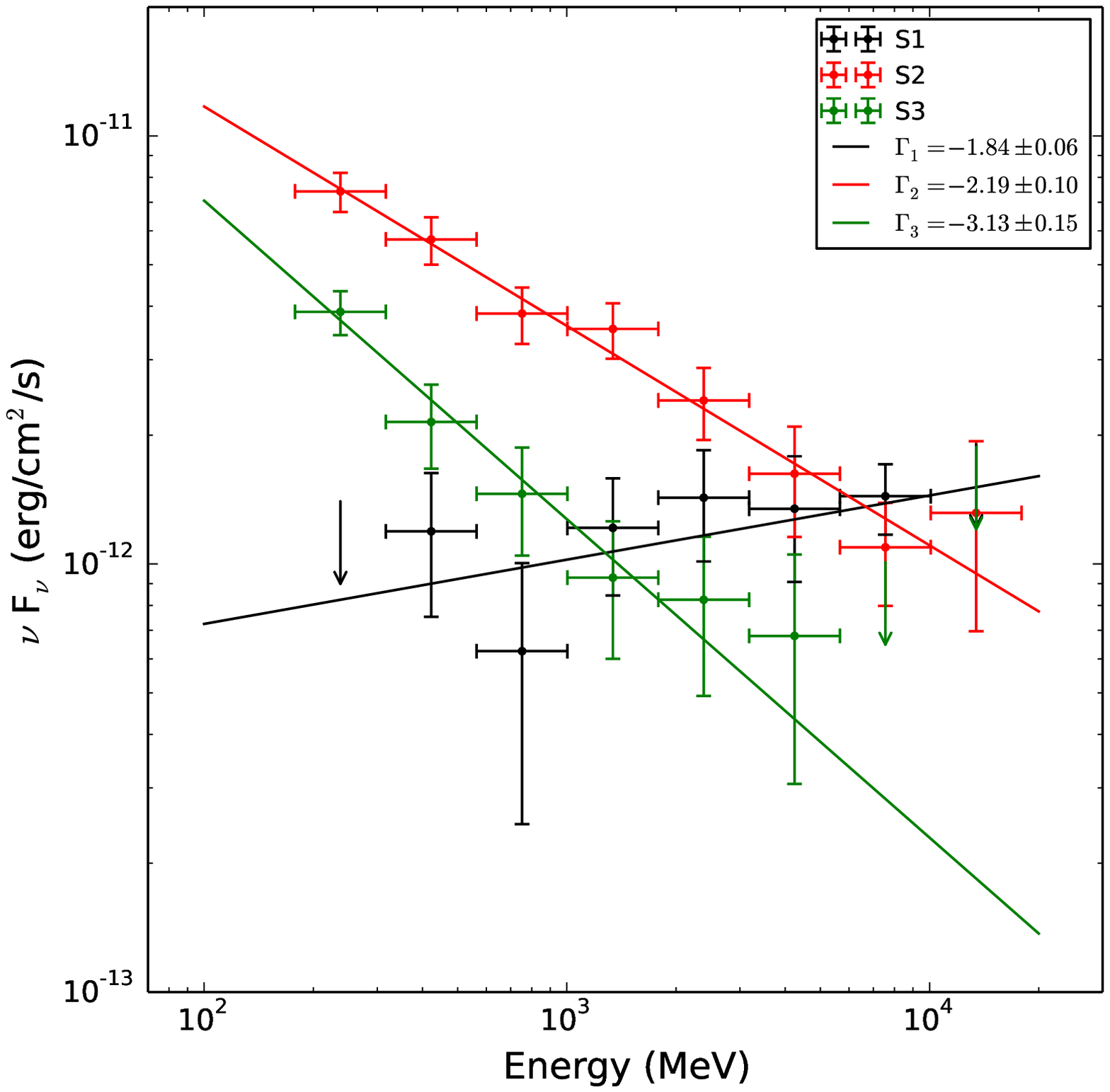}
\caption{Left plot shows the SEDs of the slices in the north lobe. The right plot shows the SEDs of the slices in the south lobe. The solid lines indicate the power-law fits. $\Gamma_1$, $\Gamma_2$, and $\Gamma_3$ are the corresponding photon indices.}
\label{fig:4}
\end{figure*}

We derived the SEDs (see Figure~\ref{fig:4}) for every slice and an upper limit was calculated within $3\sigma$ confidence level for the signal that was detected with a significance of less than $2\sigma$. We used a power-law function to fit the observed data. The upper limits in SEDs were also used to constrain the parameters of the power-law function. As shown in Figure~\ref{fig:4}, in the north  lobe N1 and N2 are consistent with each other within the uncertainties, but N3 is steeper than N1 and N2.
In the south lobe, the  photon indices for the three slices differ;  the spectra become harder moving away from the core. 

\begin{table}
\caption{TS value and likelihood value for the three templates used in \ref{sec:spatial_analy}.}      
\label{table:1}      
\centering                                      
\begin{tabular}{ccccc}          
\hline\hline                        
Model & Core & North lobe & South lobe & -log(Likelihood) \\    
\hline                                  
T1&7147&459&1591&42104\\
T2&6195&377&1610&42156\\
T3&5881&337&1566&42184\\
\hline                                             
\end{tabular}
\end{table}

\section{Radio and \planck data reduction}\label{sec:planck_analy}
\subsection{Radio data}
We used 118 MHz MWA data \citep{2013MNRAS.436.1286M}, 408 MHz Haslem data \citep{1982A&AS...47....1H}, and 1400 MHz Parkes data  \citep{2013ApJ...764..162O}. The flux densities were measured using aperture photometry over the same subregions as the \planck data (red rectangles shown in Figure~\ref{fig:3}). We used a  ds9 plug-in {\it radio flux measurement}\footnote{\url{http://www.extragalactic.info/~mjh/radio-flux.html}} to measure the flux densities for each region and frequency. The results are listed in Table~3.
%

\begin{table*}
\caption{\planck characteristics of full mission maps.}             
\label{table:2}     
\centering                                
\begin{tabular}{ccccccc}         
\hline\hline                        
Frequency & Beam FWHM & Calibration error & Systematic error & CIB correction  & Zodiacal light correction & Units factor \\ 

[GHz] & [arcmin] & $[\%]$ & [$\mu$K$_{\rm{CMB}}$] & [MJy/sr] & [MJy/sr] & [Jy/pix] \\
\hline  
30&32.29&0.35&0.19&-&-&107.90\\
44&27.00&0.26&0.39&-&-&226.02\\
70&13.21&0.20&0.40&-&-&530.61\\
100&9.68&0.09&-&0.003&1.03e-4&953.87\\
143&7.30&0.07&-&0.0079&3.57e-4&1518.00\\
217&5.02&0.16&-&0.033&1.84e-3&1933.37\\
353&4.94&0.78&-&0.13&0.01&1187.18\\
545&4.83&5.00&-&0.35&0.04&3.99\\
857&4.64&5.00&-&0.64&0.12&3.99\\
\hline                                         
\end{tabular}
\tablefoot{
The beams and the values used for \planck original data corrections are all taken from \citet{2015arXiv150201582P}.}
\end{table*}

%
\subsection{\planck data}
\planck \footnote{\url{http://www.esa.int/Planck}} \citep{2010A&A...520A...1T, 2011A&A...536A...1P}, the third generation space mission following COBE and WMAP,
was designed to measure the anisotropy of the CMB. It carries two scientific instruments and scans the sky in nine frequency bands with high sensitivity and angular resolution from $33'$ to $5'$. The Low-Frequency Instrument \cite[LFI;][]{2010A&A...520A...3M, 2010A&A...520A...4B, 2011A&A...536A...3M} covers the 30, 44, and 70 GHz bands with amplifiers cooled to 20 K. The High Frequency Instrument \cite[HFI;][]{2010A&A...520A...9L, 2011A&A...536A...4P} covers the 100, 143, 217, 353, 545, and 857 GHz bands with bolometers cooled to 0.1 K. Details about the scientific operations of the Planck can be found in \citet{2014A&A...571A...1P} and \citet{2015arXiv150201582P}.

\begin{figure*}
\centering
\includegraphics[angle=0,scale=0.23]{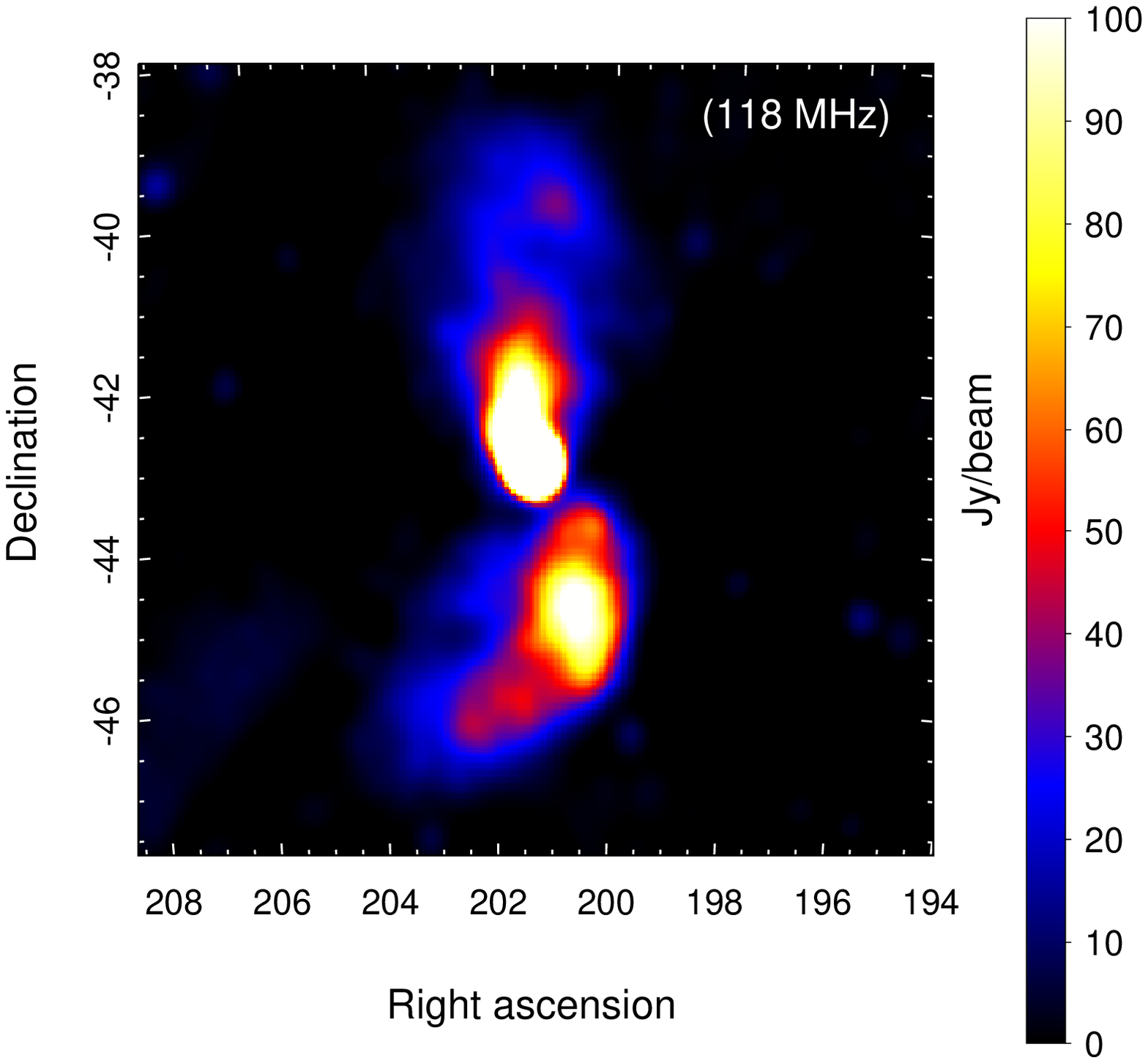}
\includegraphics[angle=0,scale=0.23]{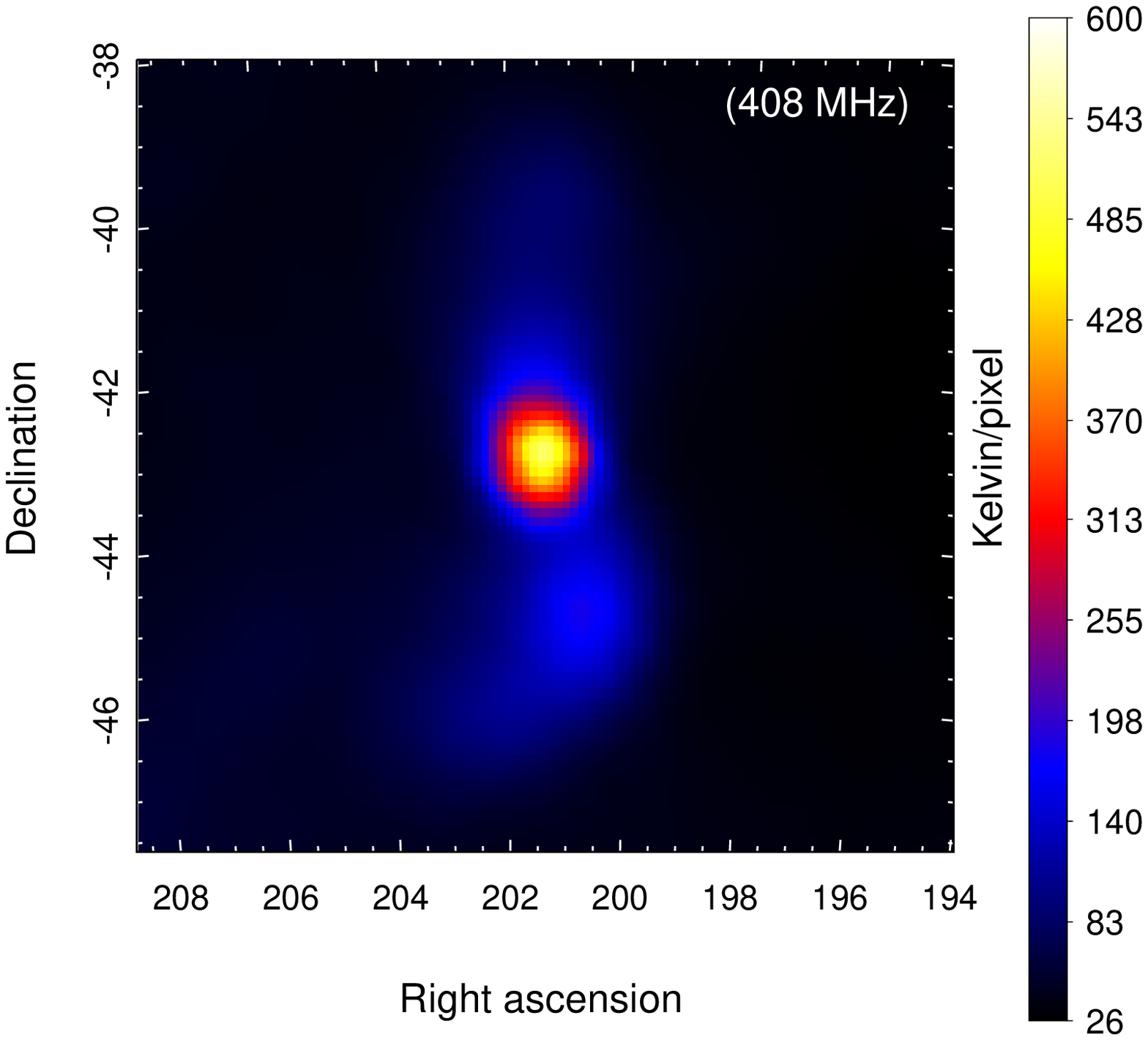}
\includegraphics[angle=0,scale=0.23]{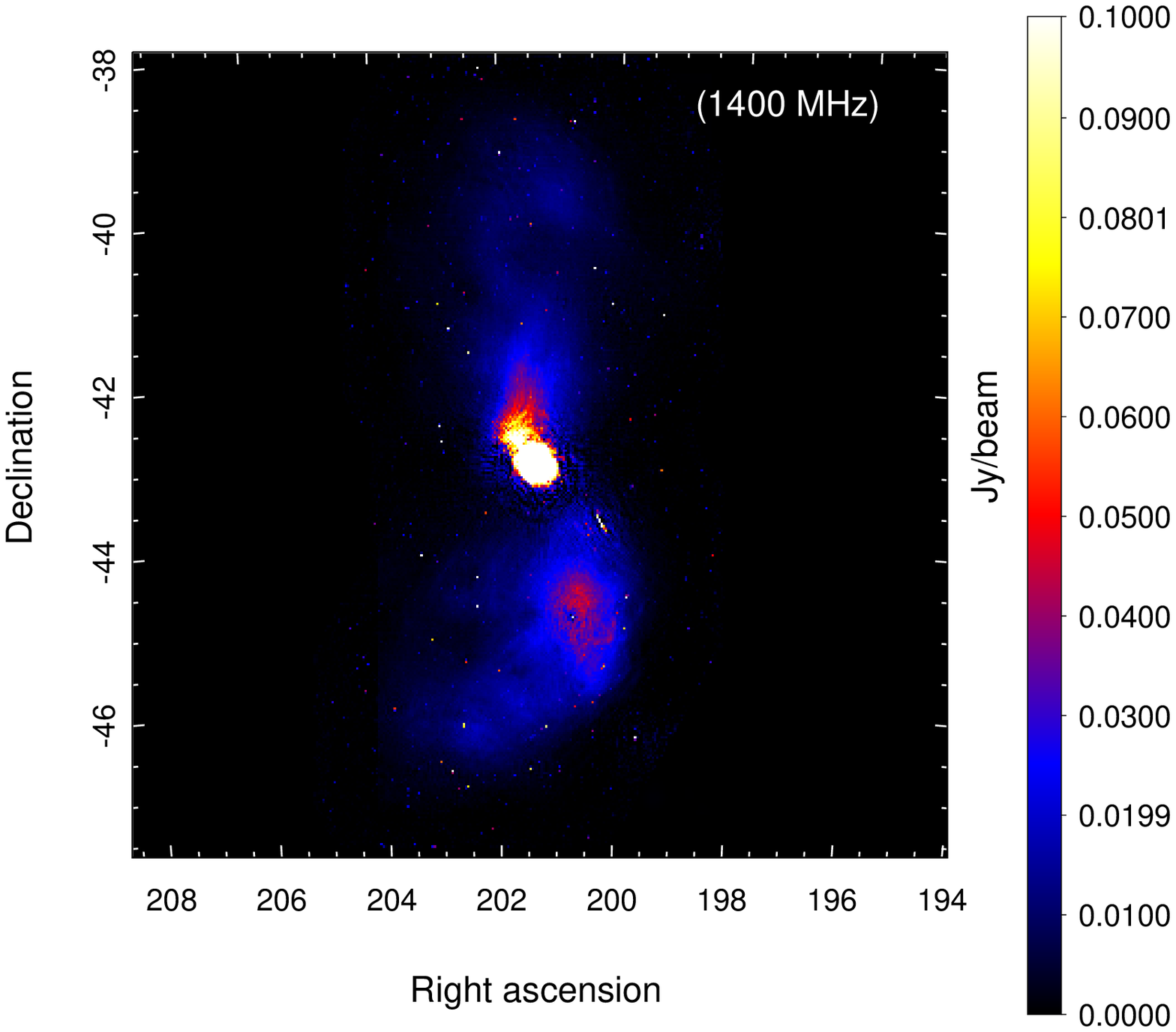}
\includegraphics[angle=0,scale=0.23]{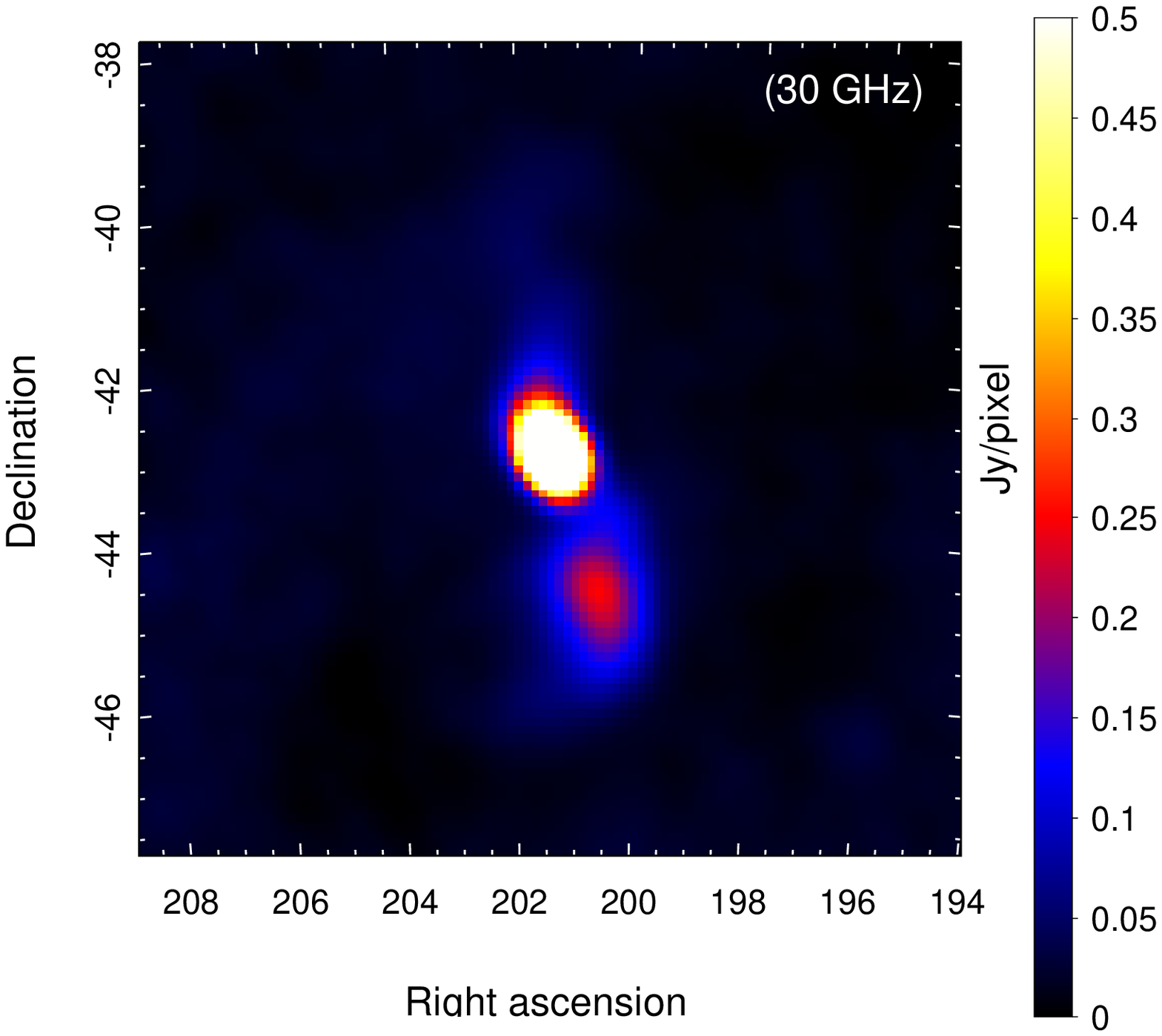}\\
\includegraphics[angle=0,scale=0.23]{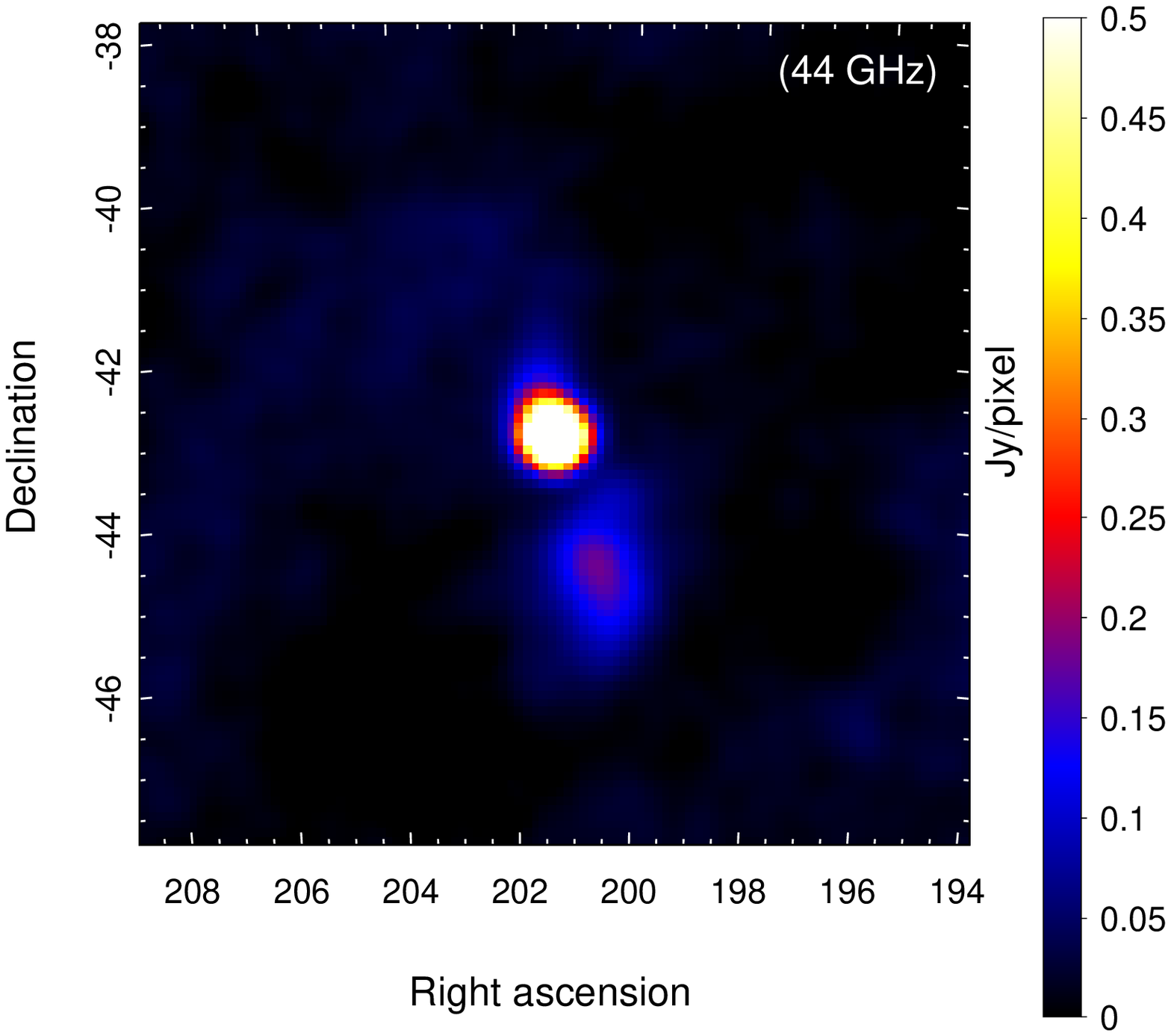}
\includegraphics[angle=0,scale=0.23]{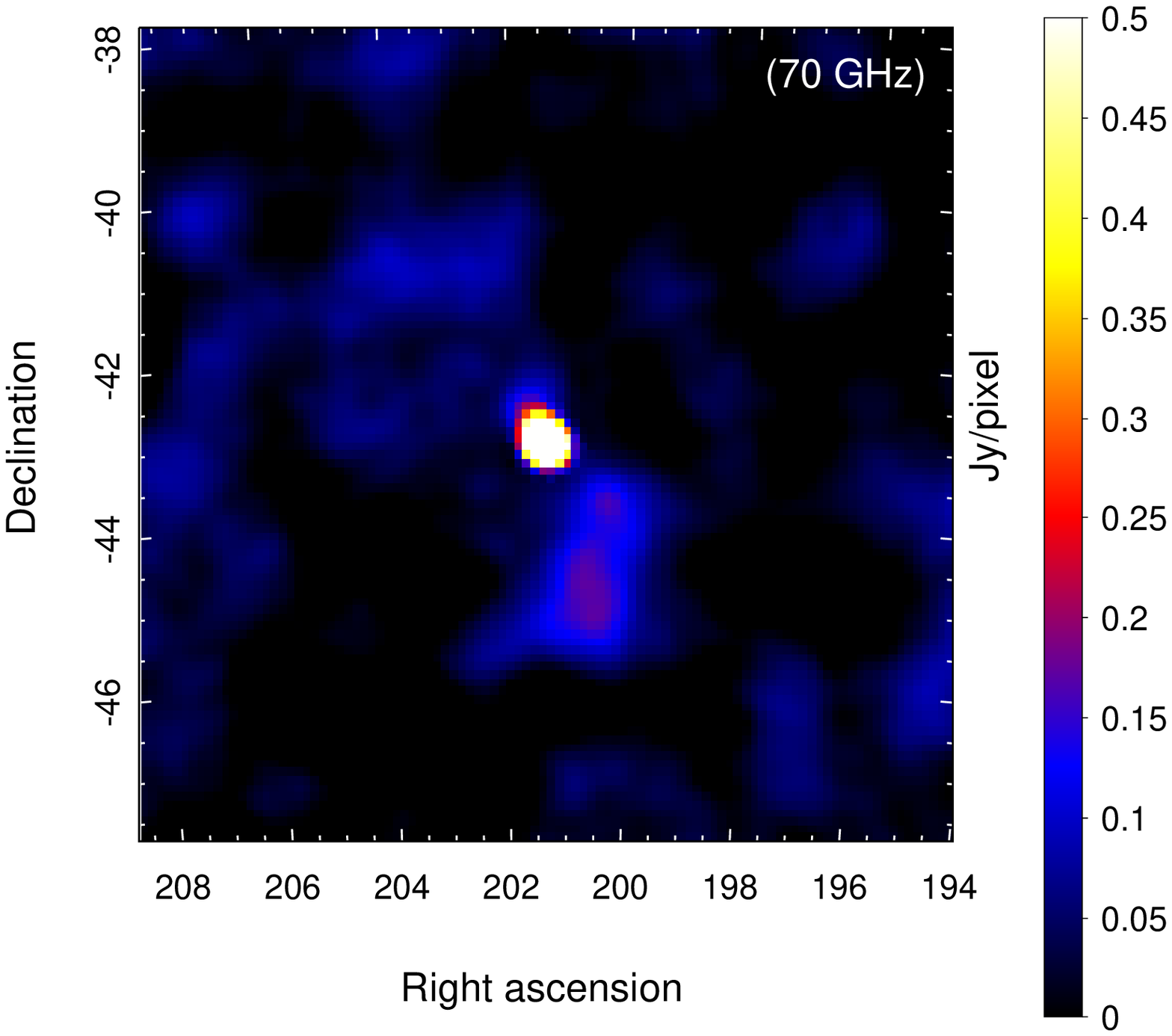}
\includegraphics[angle=0,scale=0.23]{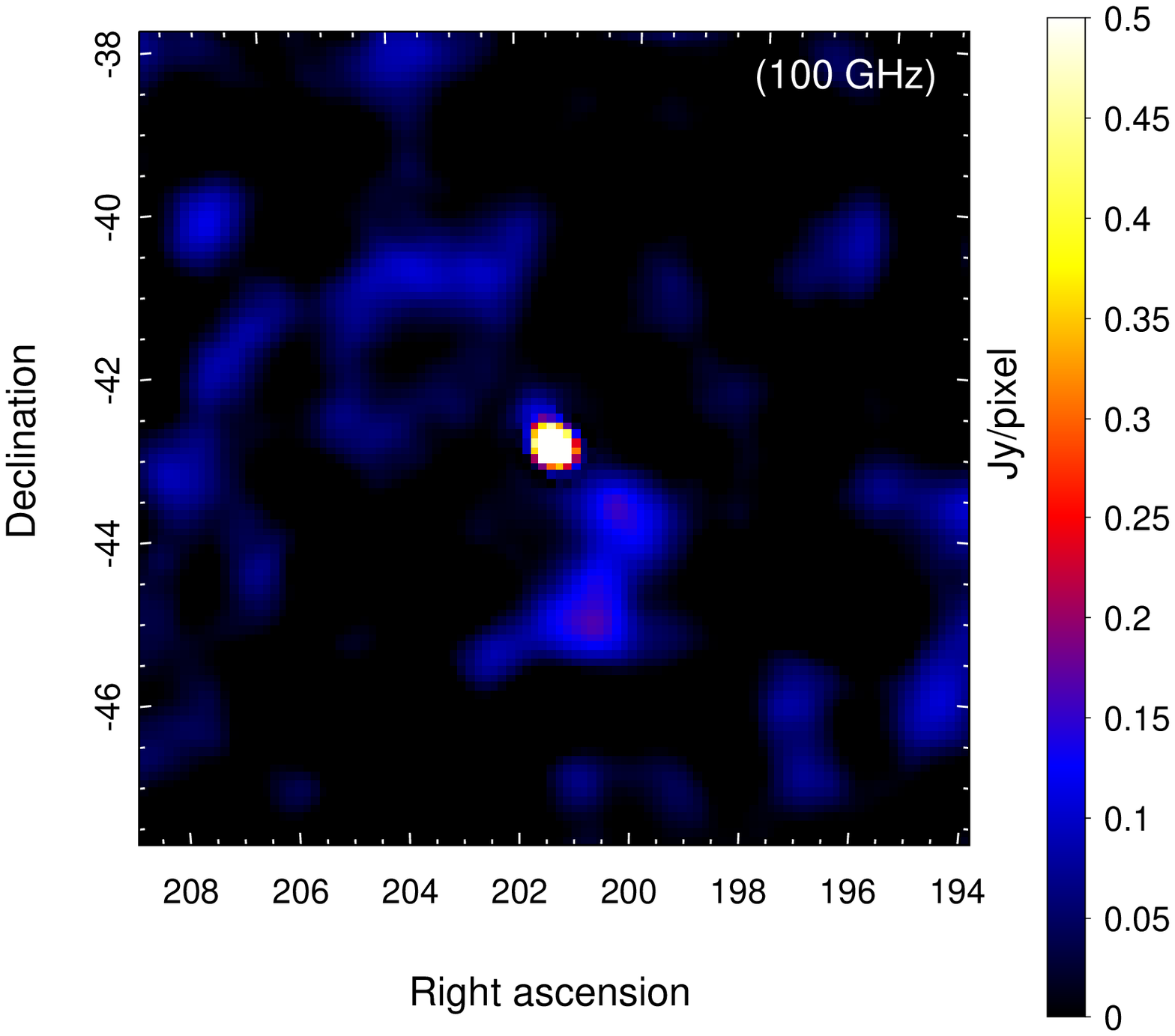}
\includegraphics[angle=0,scale=0.23]{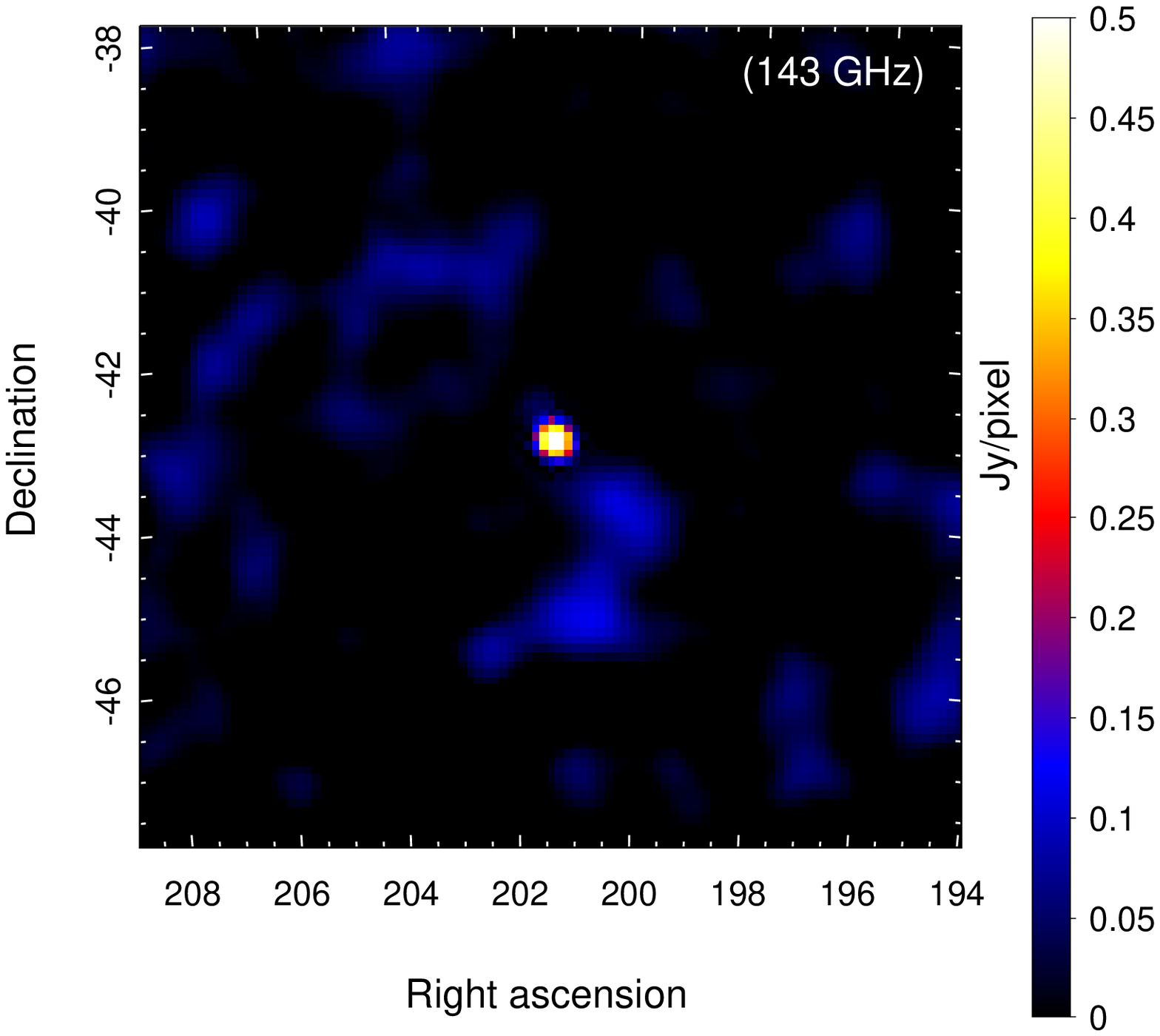}

\caption{From top left to bottom right: radio/microwave images of the Cen A and surrounding field at 118, 408, 1400 MHz, 30, 44, 70, 100, and
143 GHz. The \planck maps are {\it cleaned maps}, which are used to measure the flux densities (see subsection~\ref{sec:foreground_sep} in details). These images are smoothed using a Gaussian kernel $0.3^ \circ$.}
\label{fig:5}
\end{figure*} 

\begin{table*}
\caption{Flux density measurements of the regions of Cen A at low frequencies. Blanks indicate the signal that was not detected in that region.}             
\label{table:3}     
\centering                                
\begin{tabular}{cccccccc}         
\hline\hline                        
Frequency &   &   &   & Flux density [Jy] &  &  &  \\ 

[GHz] & N1 & N2 & N3 & core & S1 & S2 & S3 \\
\hline  
0.118\tablefootmark{a}&362.65$\pm$48.26&517.63$\pm$61.86&620.58$\pm$70.84&2877.38$\pm$295.91&479.99$\pm$60.31&849.74$\pm$95.37&764.29$\pm$86.30\\
0.408\tablefootmark{a}&165.07$\pm$18.78&265.98$\pm$28.95&291.83$\pm$31.21&1103.06$\pm$112.19&368.12$\pm$39.64&495.86$\pm$52.01&382.24$\pm$40.48\\
1.4\tablefootmark{b}&72.25$\pm$1.55&97.56$\pm$2.04&107.22$\pm$2.22&483.45$\pm$9.74&81.13$\pm$1.73&163.92$\pm$3.37&162.55$\pm$3.34\\
30&9.63$\pm$0.01&11.19$\pm$0.01&12.41$\pm$0.01&82.48$\pm$0.01&9.29$\pm$0.01&19.26$\pm$0.01&24.36$\pm$0.01\\
44&5.43$\pm$0.03&8.03$\pm$0.03&8.67$\pm$0.02&55.09$\pm$0.02&-&10.82$\pm$0.03&16.63$\pm$0.02\\
70&2.37$\pm$0.07&11.70$\pm$0.06&7.92$\pm$0.05&39.89$\pm$0.05&-&15.67$\pm$0.06&12.58$\pm$0.05\\
100&-&4.45$\pm$0.05&-&23.21$\pm$0.03&-&9.11$\pm$0.04&-\\
143&-&-&-&9.16$\pm$0.04&-&3.92$\pm$0.04&-\\
\hline               
\end{tabular}
\tablefoot{\\
\tablefoottext{a}{10\% of the flux is considered as systematic error.}
\tablefoottext{b}{2\% of the flux is considered as systematic error.}
}
\end{table*}

%
In this work we use the  \planck full-mission maps from Public Data Release 2 (PR2) products, which can be obtained via the \planck Legacy Archive (PLA) interface\footnote{\url{http://pla.esac.esa.int/pla}}.
The \planck all-sky maps are in Healpix \citep{2005ApJ...622..759G} format, with the resolution parameter $N_{\rm side} = 1024$ for LFI 30, 44 and 70 GHz, and 2048 for LFI 70 GHz and HFI $100-857~\rm GHz$. The data are given in units of CMB thermodynamic temperature (K$_{\rm CMB}$) up to $353~\rm GHz$, and in $\rm MJy~sr^{-1}$ for $545$ and $857~\rm GHz$. The temperature units and $\rm MJy~sr^{-1}$ were converted to flux density per pixel with multiplication by the factor given in the last column of Table~2. 
For ease of comparison with \gray emission, we degraded the resolution of the original Healpix data from 1024 or 2048 into 512 (a pixel size of about $6'$), and then projected them on to the area with \fermi's sky map using the Healpy package$\footnote{\url{https://healpy.readthedocs.org/en/latest/tutorial.html}}$ and Astropy package$\footnote{\url{http://docs.astropy.org/en/stable/index.html#}}$. %
Both calibration and systematic uncertainties were considered. The zodiacal light level corrections were added to the maps, and the cosmic infrared backgrounds (CIB) were removed from the maps \citep{2015arXiv150201582P}. 
The characteristics of \planck for each frequency band are listed in Table~2.
%

\subsection{Thermal dust and CMB components separation}\label{sec:foreground_sep}
In our ROI, synchrotron radiation dominates in the  low frequency band. However, in the high frequency bands ($> 100$ GHz)  CMB and thermal dust emission begin to overwhelm the non-thermal emission \citep{2015arXiv150201588P}. It is possible to use higher frequency maps to derive accurate information on CMB anisotropy and thermal dust emission, with which we can get more robust non-thermal spectra in lower frequency than a simple aperture photometry method.   

Thermal emission from dust grains dominates the radiation mechanism in the far-infrared (FIR) to millimetre range (e.g. \citealt{, 2003ARA&A..41..241D,2007ApJ...657..810D,2011A&A...525A.103C}), and is  the main foreground hampering the study of  Cen A at \planck frequencies. In this paper we selected a modified blackbody (MBB) \citep{2014A&A...571A..11P} to fit the thermal dust component empirically, that is

\begin{equation}
\label{equation:TD_function}
I_{\rm d} = A_{\rm d}B_\nu\left(T_{\rm obs}\right)\left(\frac{\nu}{\nu_0}\right)^{\beta_{\rm obs}}, 
\end{equation}
where $\nu_0 = 353~{\rm GHz}$ is the reference frequency.
There are three parameters in this model: the dimensionless amplitude parameter $A_{\rm d}$, temperature $T_{\rm obs}$, and the spectral index $\beta_{\rm obs}$. Because there are only a few frequency bands available, in the fitting we leave $A_{\rm d}$ and $T_{\rm obs}$ to be free and fix the index $\beta_{\rm obs}$ to be the value in the  Planck all-sky model of thermal dust emission \citep{2014A&A...571A..11P} from PLA.

Another foreground is the CMB. A blackbody with the $T_{\rm CMB} = 2.7255$ K \citep{2009ApJ...707..916F} is selected to fit the CMB component

\begin{equation}
\label{equation:CMB_function}
I_{\rm CMB} = A_{\rm CMB}B_\nu\left(T_{\rm CMB}\right),
\end{equation}
where $A_{\rm CMB}$ is a dimensionless amplitude parameter.

In order to derive the emission signals of Cen A in the low energy band from 30 to 143 GHz accurately, we used the \planck 217, 353, 545, and 857 GHz, and IRAS 3000 GHz ($100~\mu m$) data  to constrain the parameters $A_{\rm d}$, $T_{\rm obs}$, and $A_{\rm CMB}$ at each pixel using a chi-squared ($\chi^2$) minimisation, and then extrapolated the two models to low energy. Here we manipulated the $100~\mu m$ data following the method described in \citet{2014A&A...571A..11P}. 

Considering the effects of both the parameter uncertainties of the models (CMB and thermal dust) and the errors of the observed \planck data, we performed the following steps in deriving the microwave flux and errors of the lobes. (1) For any pixel within the ROI, we drew 50 groups  of random samples from a normal (Gaussian) distribution for the free parameter set ($A_{\rm CMB}$, $A_{\rm d}$, and $T_{\rm obs}$) according to the best-fitted value and fitted uncertainties. The same procedure was also applied to the observed \planck data in each pixel. The observed uncertainties should obey a Poisson distribution rather than a Gaussian, but the large counts of the Planck maps makes Gaussian statistics a reasonable assumption. (2) For each group of samples, we calculated the thermal dust and CMB components based on  equation~\ref{equation:TD_function} and equation~\ref{equation:CMB_function}, respectively. The thermal dust and CMB maps were both smoothed to the \planck original angular resolution (listed in Table~2) to obtain a matched resolution map. Then we removed the CMB and thermal dust component to derive the {\it background subtracted} value of this pixel in each sample. 
(3) We chose the average and standard deviation of the 50 sampled {\it background subtracted} values as the final {\it cleaned} value and corresponding errors in this pixel.   (4) We repeated steps (1) to (3) at each pixel within the ROI, and finally derived the CMB and dust emission subtracted {\it cleaned} maps and corresponding error maps, which were used to measure the integral flux densities in the following. The derived {\it cleaned} maps from 30 GHz to 143 GHz are shown in Figure~\ref{fig:5}.
%

\subsection{Flux density measurements}

We used the standard aperture photometry, with the aperture size (red rectangles) shown in Figure~\ref{fig:3}, to measure the integral flux densities of the \planck 30, 44, 70, 100, and 143 GHz maps. The measurements of flux density for each region and frequency, together with errors within $1\sigma$ confidence level, are listed in Table~3. The errors were derived from the error maps described above  using  error propagation. We can see the flux densities of the core are consistent with their corresponding values of region 3 in \citet{2009MNRAS.393.1041H}, which confirm that our model selection and method are reasonable.
%

\begin{figure}
\centering
\includegraphics[angle=0,scale=0.6]{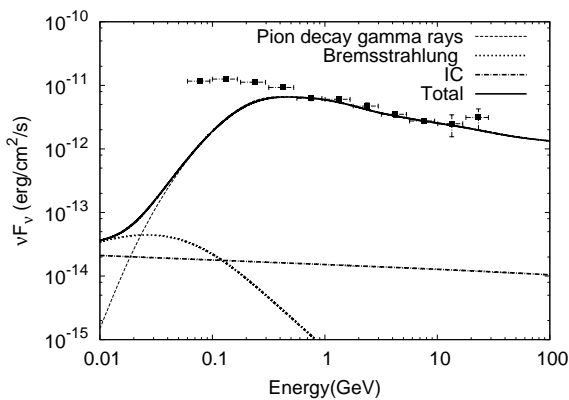}

\caption{$\gamma$-rays from hadronic interactions of cosmic rays in different channels.  Also shown is the SED of the south lobe. The primary proton spectra are assumed to be a power-law function with an index of 2.3 to fit the high energy part of the SED of the south lobe.   }

\label{fig:spec}
\end{figure}

\begin{figure*}
\centering
\includegraphics[angle=0,scale=0.29]{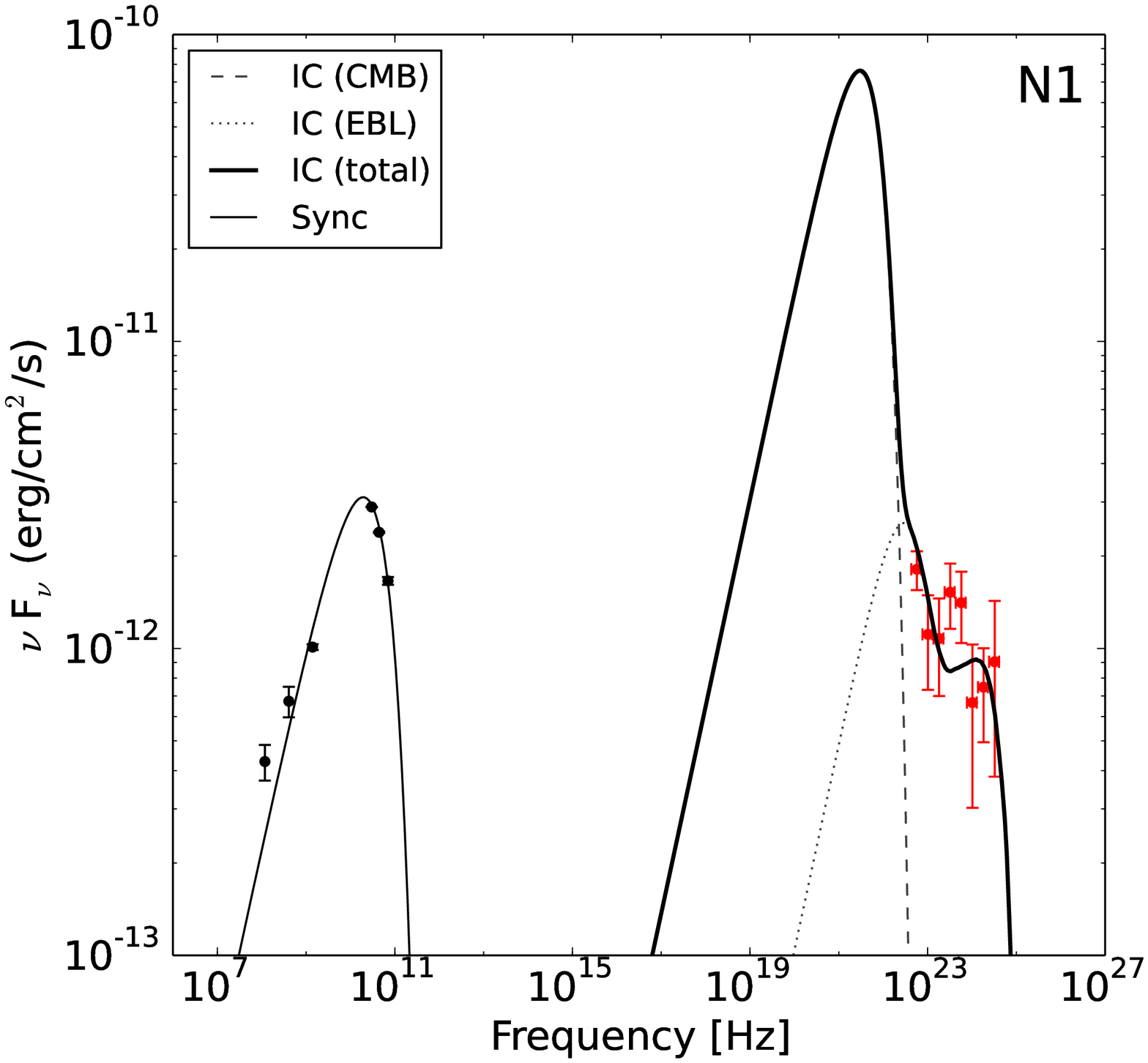}
\includegraphics[angle=0,scale=0.29]{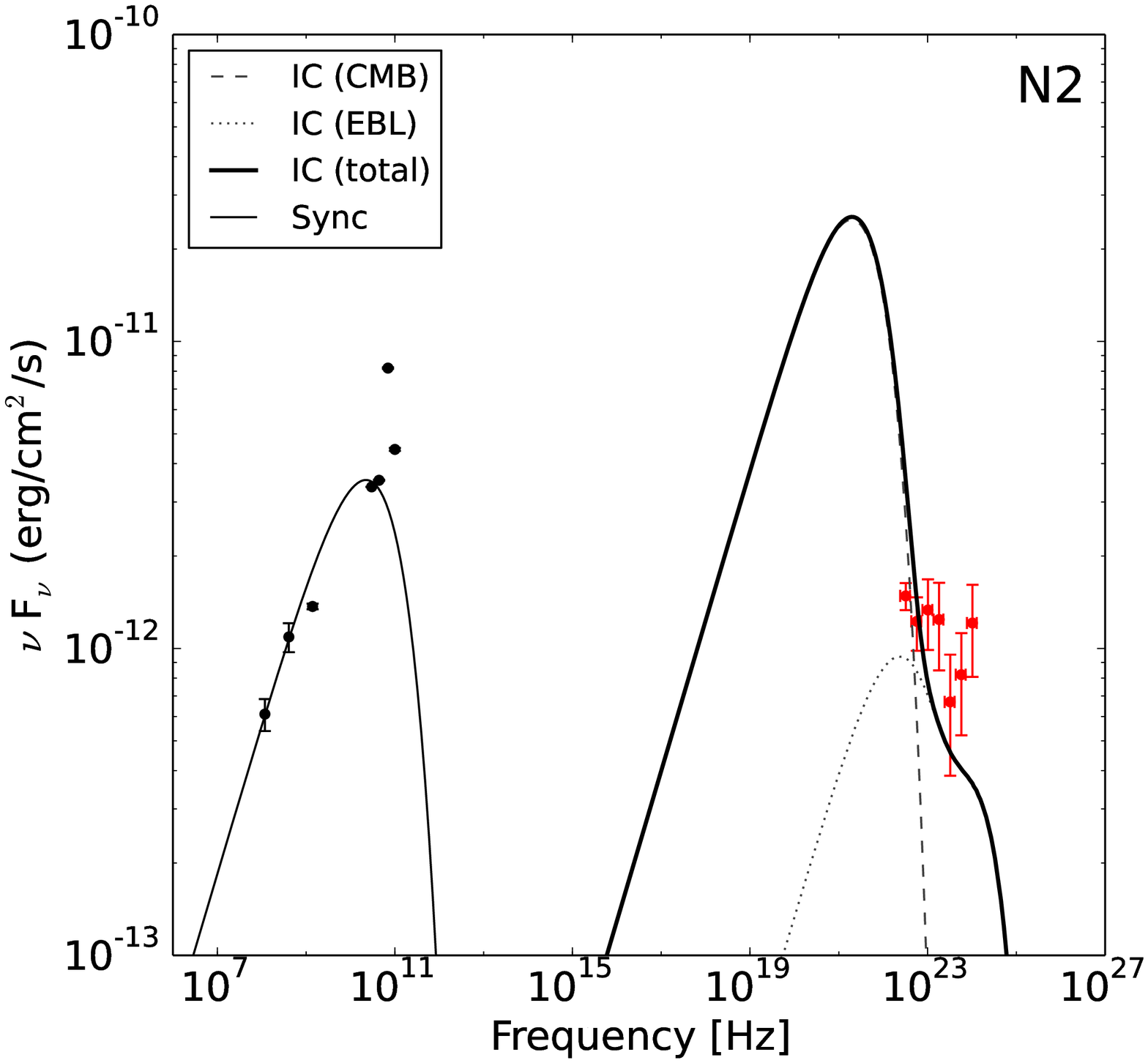}
\includegraphics[angle=0,scale=0.29]{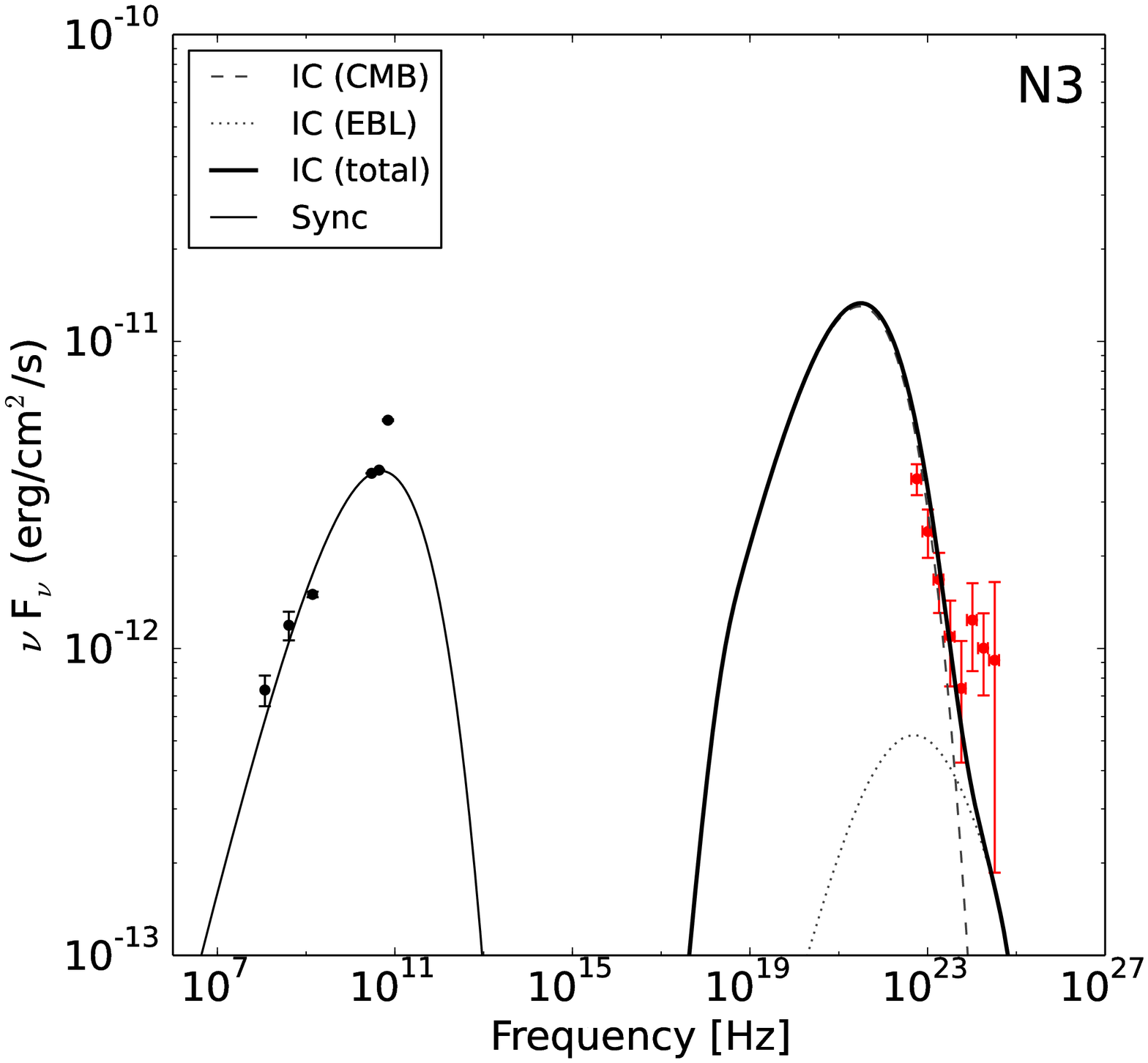}\\
\includegraphics[angle=0,scale=0.29]{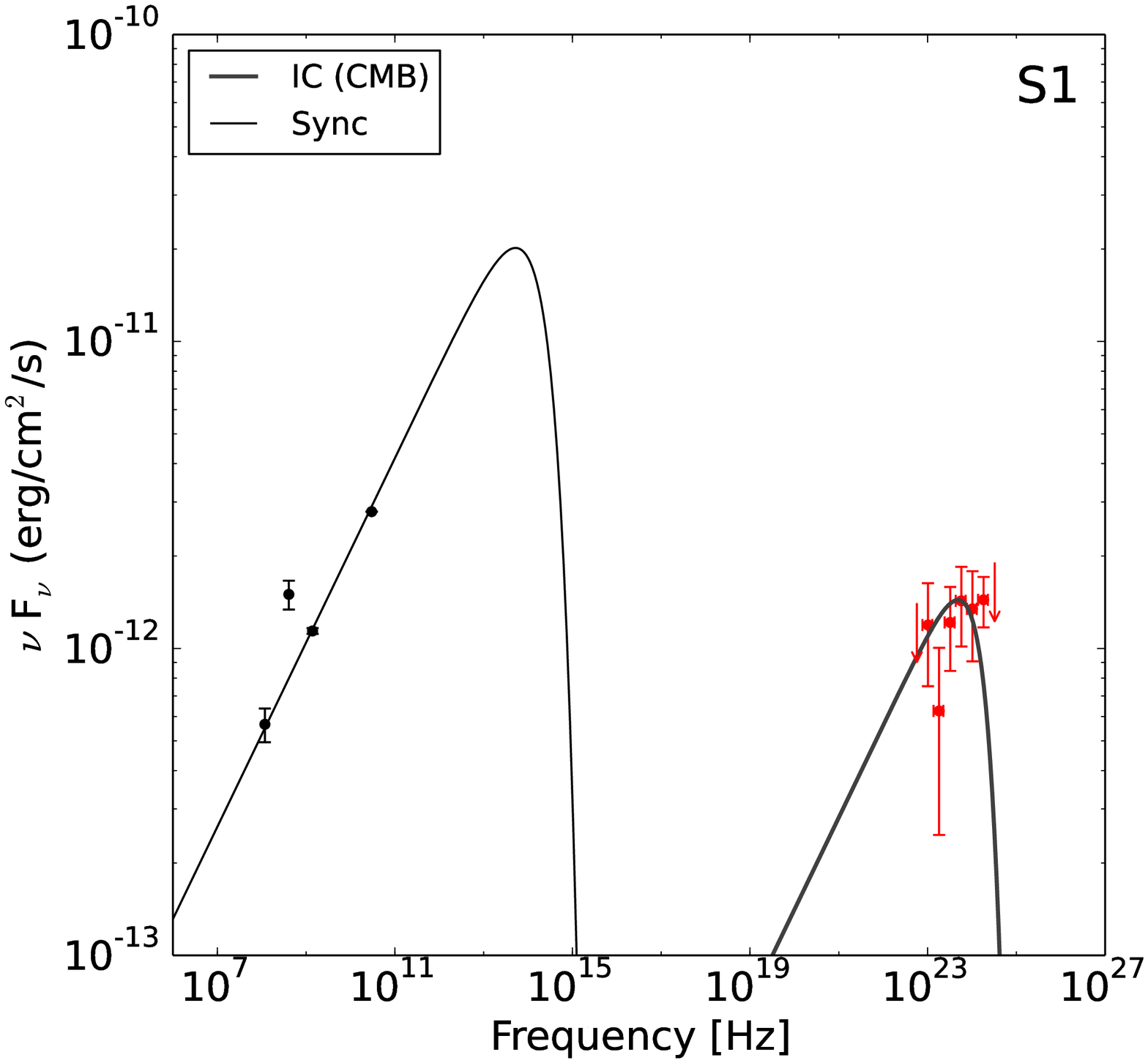}
\includegraphics[angle=0,scale=0.29]{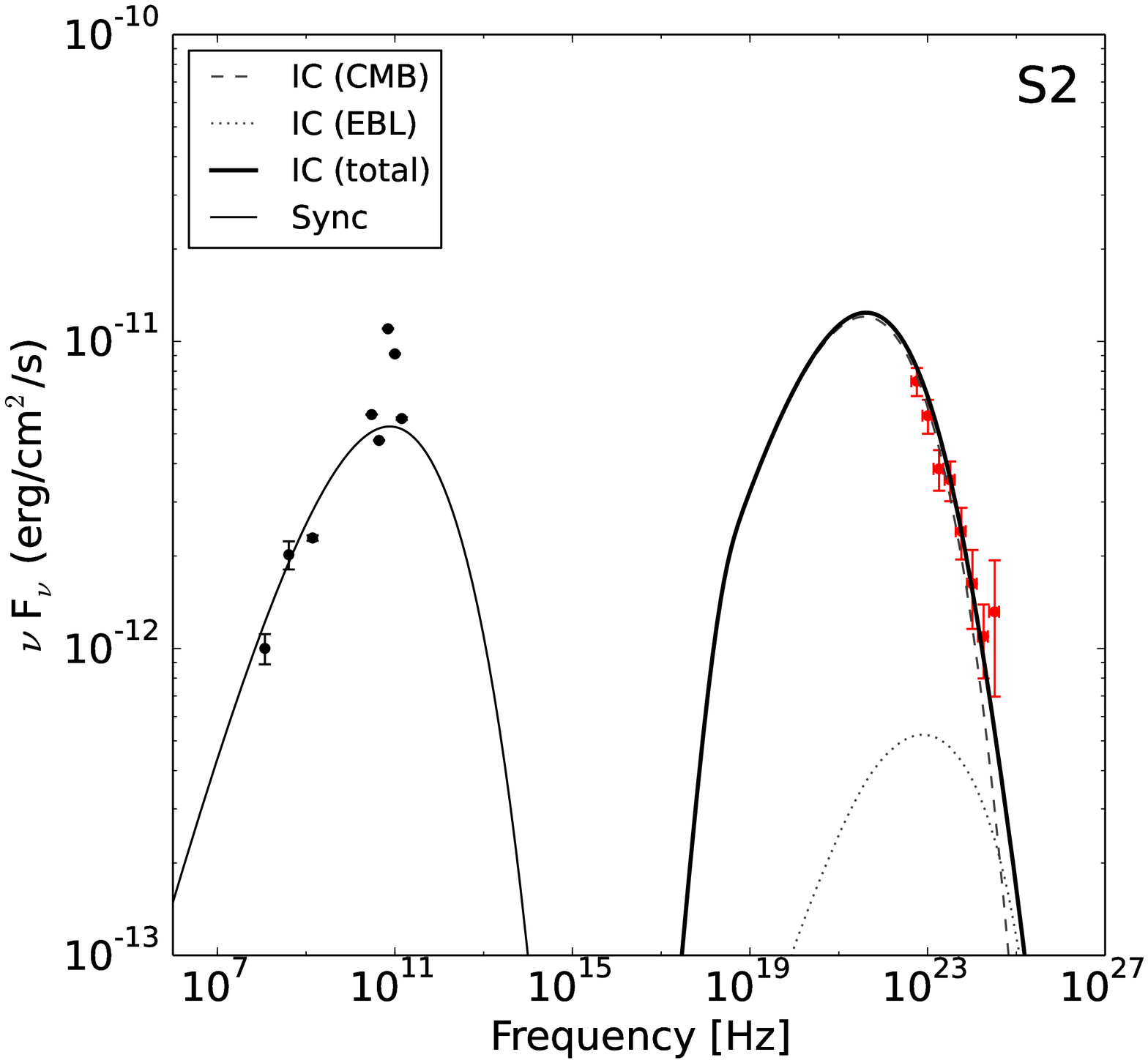}
\includegraphics[angle=0,scale=0.29]{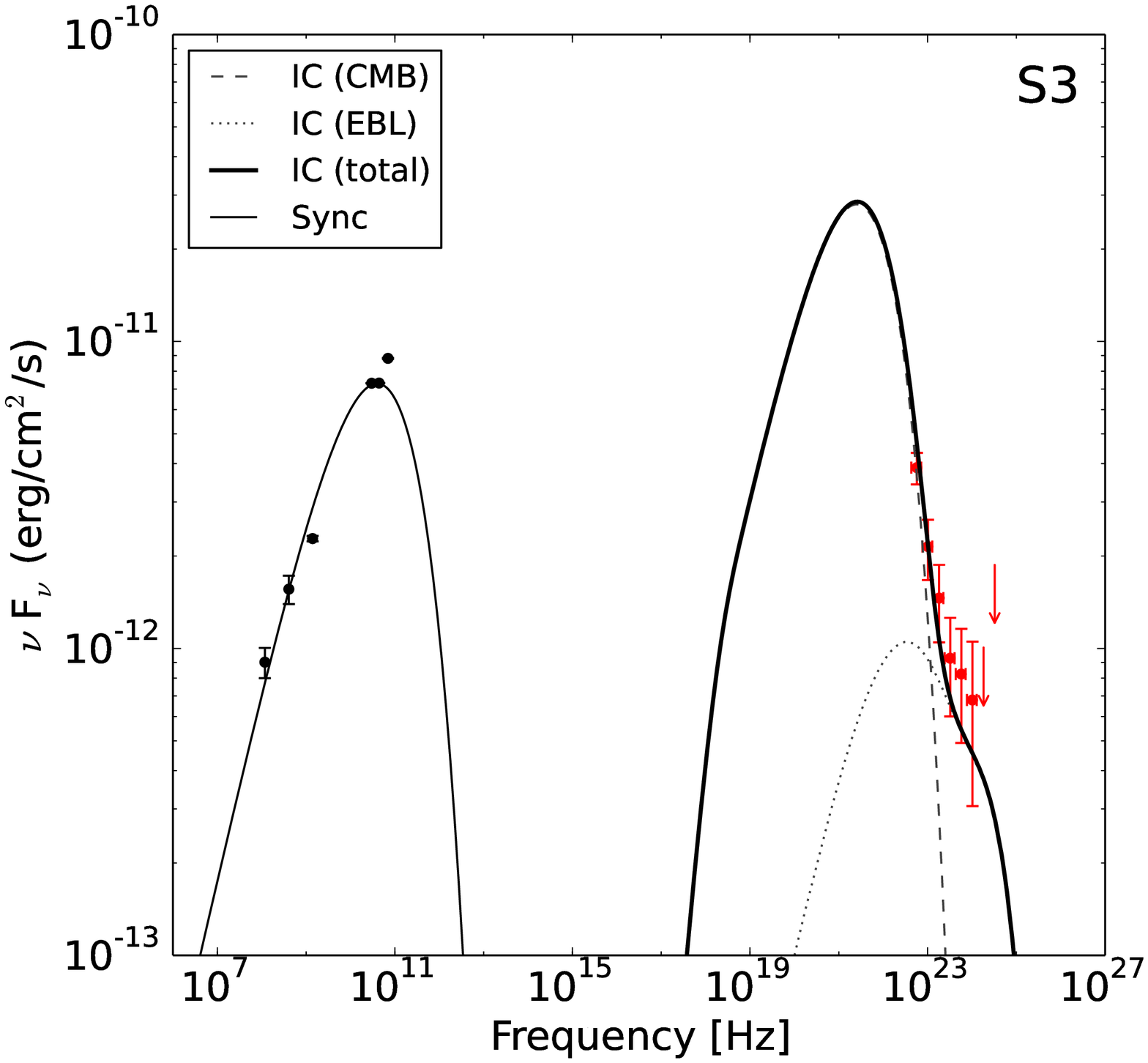}
\caption{Broadband SEDs for each region shown in Figure~\ref{fig:3}. Observed radio and \planck data (black dots with error bars) are fitted with a synchrotron model. Observed \fermi data (red dots with error bars) are fitted with the inverse-Compton (IC) scatterings of the CMB and EBL photon fields except for S1, which only requires the seed photon contribution from the CMB. The upper limits are calculated within a $3\sigma$ confidence level.}
\label{fig:SED_fitting}
\end{figure*}

\begin{table*}
\centering 
\caption{Summary of SED best-fitting model parameters for the  power-law electron distribution with cutoff.}             
\label{table:4}
\begin{tabular}{ccccccc}
\hline\hline
Model components & N1 & N2 & N3 & S1 & S2 & S3 \\
\hline
$W_{\rm e}$~[$\times 10^{57}$ erg]&0.98$\pm$0.08&1.8$\pm$0.3&0.67$^{+0.07}_{-0.05}$&0.18$\pm$0.03&2.5$\pm$1.5&1.0$^{+0.3}_{-0.2}$\\
$\alpha$&1.65$\pm$0.03&2.02$^{+0.06}_{-0.07}$&1.79$\pm$0.04&2.42$\pm$0.02&1.96$^{+0.07}_{-0.14}$&1.78$\pm$0.07\\
$E_{\rm cutoff}$ [GeV]&65.5$\pm$1.8&59$\pm$6&21.2$\pm$1.9&-&5.0$\pm$0.4&38.9$^{+3.0}_{-1.8}$\\
$\beta$&20$\pm$2&1.9$\pm$0.2&0.68$\pm$0.02&-&0.42$^{+0.05}_{-0.02}$&1.02$^{+0.07}_{-0.05}$\\
B [$\mu \rm G$]&0.70$\pm$0.04&1.26$\pm$0.08&1.78$^{+0.07}_{-0.11}$&13.4$\pm$0.8&2.29$^{+0.17}_{-0.12}$&1.74$\pm$0.07\\  
\hline
\end{tabular}
\end{table*}

\section{Modelling the spectral energy distributions}\label{sec:fitting}
To fit the derived spectral distributions, we used the software package {\it Naima}\footnote{\url{http://naima.readthedocs.org/en/latest/index.html#}}.  We found that a power-law shape of the spectrum is extended down to  100 MeV. This  does not agree with the spectrum of  low energy 
$\gamma$-rays from hadronic interactions of cosmic rays. The latter has a  standard shape, which is dictated by the kinematics of the $\pi^0$ decay rather than the spectrum of cosmic rays.   The region of sharp decline of the 
SED  of $\pi^0$-decay  $\gamma$-rays below 1 GeV is partly "filled"  by photons generated by secondary electrons (through the IC  scattering and Bremsstrahlung), that is the products of the charged  $\pi^\pm$-meson decays. 
However, these components do not appear to sufficiently compensate the deficit even in the most optimistic case of  ``thick target" when the production of secondaries is saturated. This is demonstrated in Fig.\ref{fig:spec}, which shows the three channels of $\gamma$-ray production initiated by $pp$ interactions and the SED of south lobe. It is assumed that the density  of the  ambient gas $n$ 
is sufficiently high that the lifetimes of relativistic protons, $t_{\rm pp} \sim 10^{15} \left(\frac{n}{1~ \rm cm^{-3}}\right)^{-1} \rm s$, as well as secondary electrons,  are  shorter than the confinement time of cosmic rays. 
Under this condition, the steady-state solutions shown in Fig.\ref{fig:spec} apparently do not depend on the density $n$. The relative contribution of $\gamma$-rays from secondary electrons does not strongly depend
on the spectrum of cosmic rays. We can safely conclude that low energy $\gamma$-rays from both lobes are  not contributed by  cosmic ray protons and nuclei.

Thus, the spectral measurements presented in this paper remove the uncertainty of our previous study, \citet{2012A&A...542A..19Y}, regarding the origin of $\gamma$-rays. It is clear that $\gamma$-rays are produced, at least in the energy band below 1 GeV, by directly accelerated electrons.  Because of  the low gas density in the lobes, the $\gamma$-ray production is dominated by the IC scatterings of  photons of the 2.7 K CMB radiation, with possible  contribution from photons of the EBL by relativistic electrons.  Although the energy density of the EBL is much lower than the energy density of the CMB, the role of the EBL photons can be noticeable in the formation of the spectrum at  the highest $\gamma$-ray energies, 
especially in the case of a cutoff in the electron spectrum below  a few  TeV. The photons from the host galaxy of Cen A are also  potential seed photons for IC scattering. As calculated in \citet{2010Sci...328..725A}, however, the IC gamma rays from the photon fields produced by the host galaxy are negligible compared with those  from the CMB and EBL.
 
For different parts of the lobes, the  distributions of electrons and the strength of 
the magnetic fields can be
derived  from the fit of the  \planck  radio and  the \fermi $\gamma$-ray data by synchrotron and IC components, respectively. We used the formalism of  \citet{2010PhRvD..82d3002A} for calculations of synchrotron
radiation and the formalism  proposed in \citet{2014ApJ...783..100K} for IC scattering. The temperature $T_{\rm CMB} = 2.7255$ K and energy density $n_{\rm CMB} =0.261~\rm eV~cm^{-3}$ were adopted for the CMB photon field. We used the model of \citet{2008A&A...487..837F} for EBL.

For the energy distribution of electrons, we  assumed  the following general form:
\begin{equation}
N(E) = A~\left(\frac{E}{E_{0}}\right)^{-\alpha}~{\rm exp}(-\left(\frac{E}{E_{\rm cutoff}}\right)^{\beta}).
\end{equation}
Here $E_{0} = 1\rm~GeV$ is the reference energy. In calculations,  the parameters $A$, 
$\alpha$,  $E_{\rm cutoff}$, and $\beta$, characterising the electron spectrum, and the strength of the  magnetic field $B$, are left as free parameters. The minimum electron energy  is set to
$E_{\rm emin}=1 \rm~MeV$. %
Figure~\ref{fig:SED_fitting} shows the SED results obtained for the subregions from Figure~\ref{fig:3}. The derived model parameters $E_{\rm cutoff}$, $\beta$, B, and the corresponding errors with $1 \sigma$ confidence level, as well as the total energy of electrons $W_{\rm e}$, 
are presented in Table~4 for the regions N1,N2, N3 and S1, S2, S3.  It should be mentioned that for N2, N3, S2, and S3, the \planck data points above 70 GHz are significantly above the model predicted value and we omit these points in the fit. The reason may be a poor understanding of the high frequency background in this band.  Meanwhile, for N2 and N3 the high Fermi points are significantly above the model curve; this is caused by the fact that the weighting of these high energy points in MCMC fitting is relatively small owing to their larger error bars.  
%

\begin{table*}
\centering
\caption{Summary of SED best-fit model parameters for a broken power-law electron distribution.}
\label{table:5}
\begin{tabular}{ccccccc}
\hline\hline
Model components & N1 & N2 & N3 & S1 & S2 & S3 \\
\hline
$W_{\rm e}$~[$\times 10^{57}$ erg]&0.63$\pm$0.07&2.7$^{+0.8}_{-0.5}$&0.55$\pm$0.07&0.18$\pm$0.03&2.8$^{+2.0}_{-0.9}$&1.2$^{+0.7}_{-0.4}$\\
$\alpha_{1}$&1.54$\pm$0.06&2.13$^{+0.06}_{-0.08}$&1.92$^{+0.05}_{-0.08}$&2.42$\pm$0.02&2.08$\pm$0.10&1.77$\pm$0.14\\
$\alpha_{2}$&8.5$^{+2.0}_{-1.4}$&9$\pm$3&3.9$\pm$0.2&-&4.18$\pm$0.16&5.5$^{+0.4}_{-0.3}$\\
$E_{\rm break}$ [GeV]&51$\pm$4&59$^{+8}_{-12}$&32$\pm$3&-&36$\pm$3&44$\pm$4\\
B [$\mu \rm G$]&0.79$\pm$0.06&1.14$\pm$0.09&1.79$\pm$0.12&13.4$\pm$0.8&1.37$\pm$0.08&1.23$\pm$0.09\\
\hline
\end{tabular}
\end{table*}

\begin{figure*}
\centering
\includegraphics[angle=0,scale=0.29]{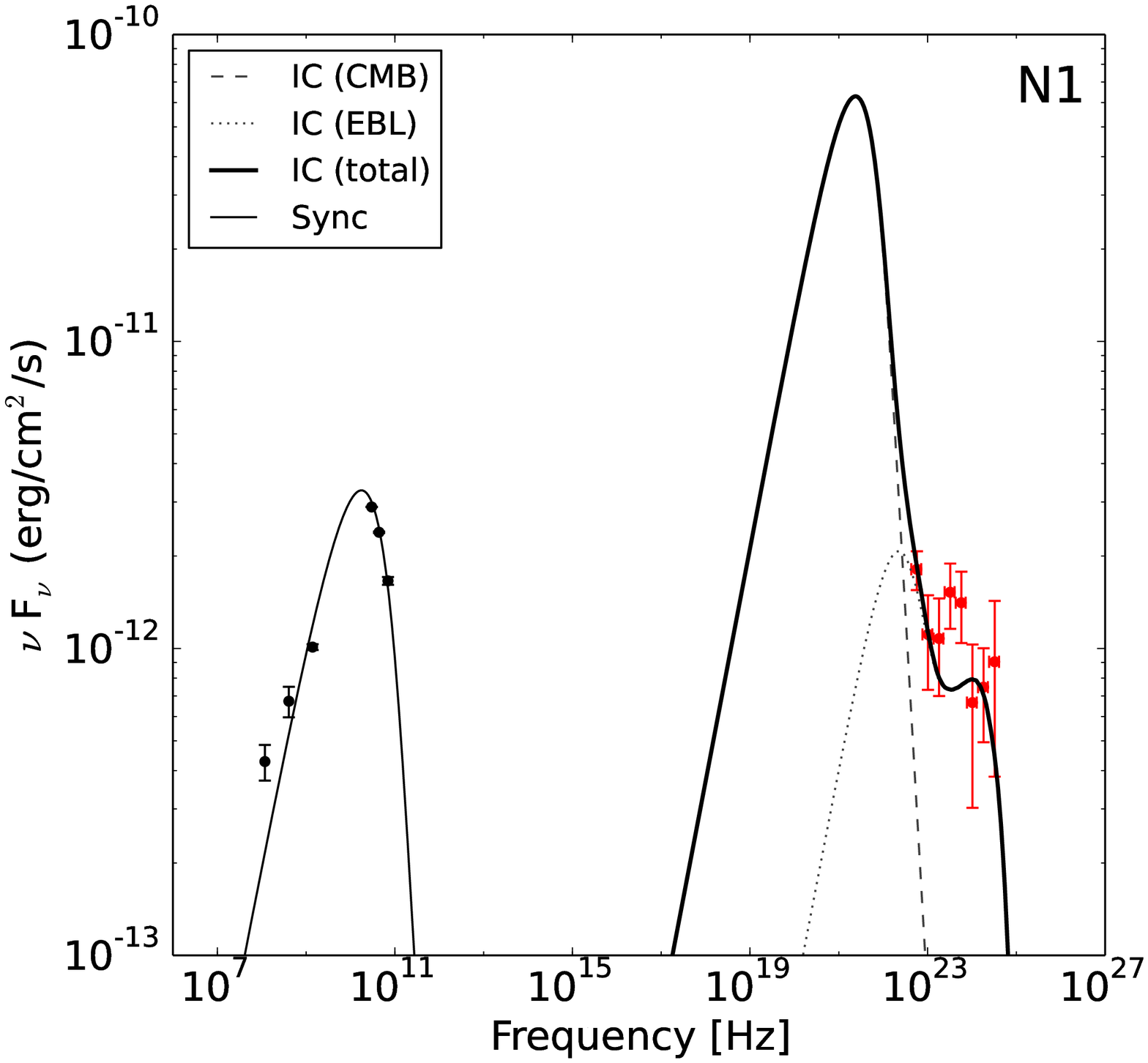}
\includegraphics[angle=0,scale=0.29]{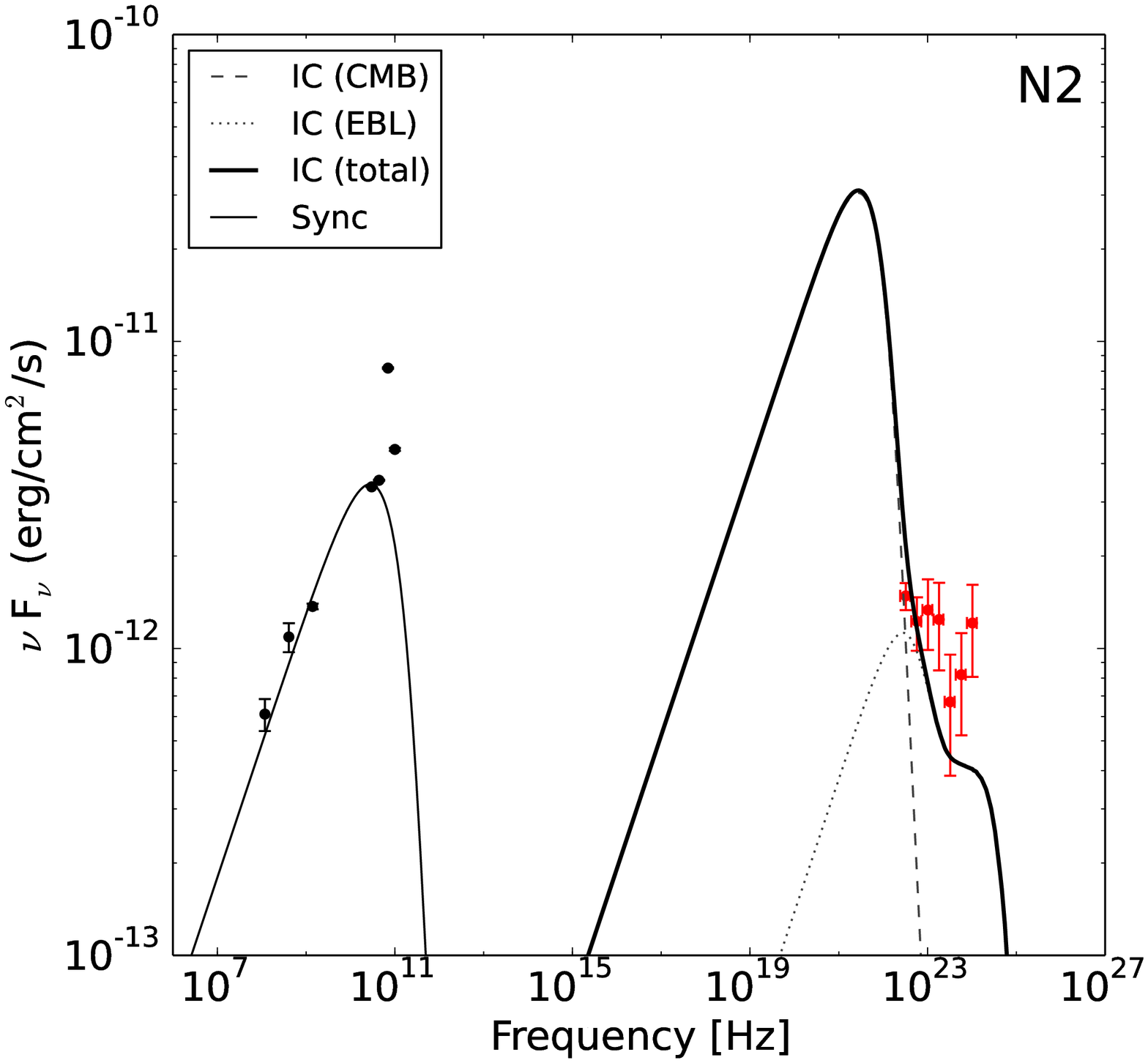}
\includegraphics[angle=0,scale=0.29]{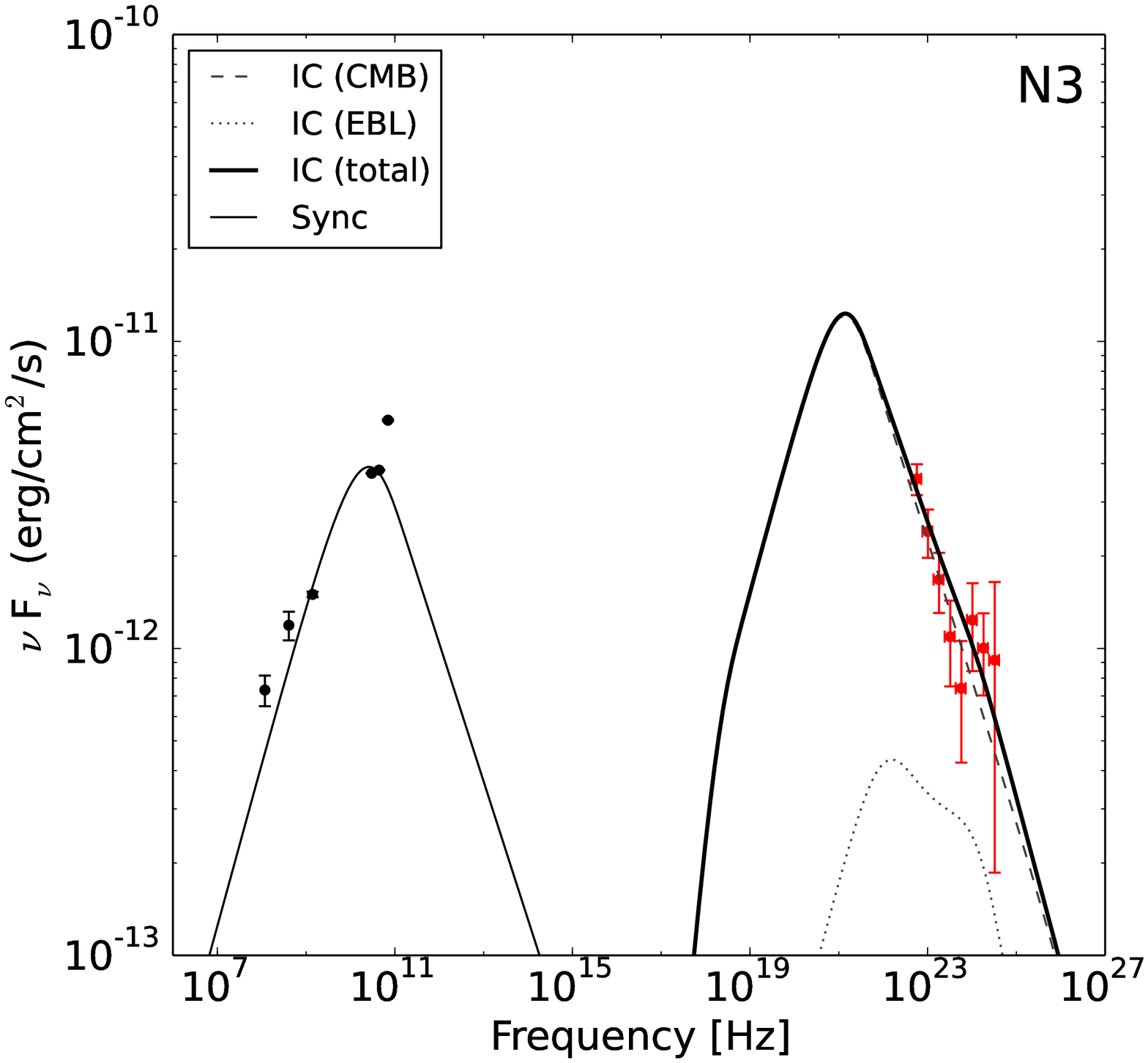}\\
\includegraphics[angle=0,scale=0.29]{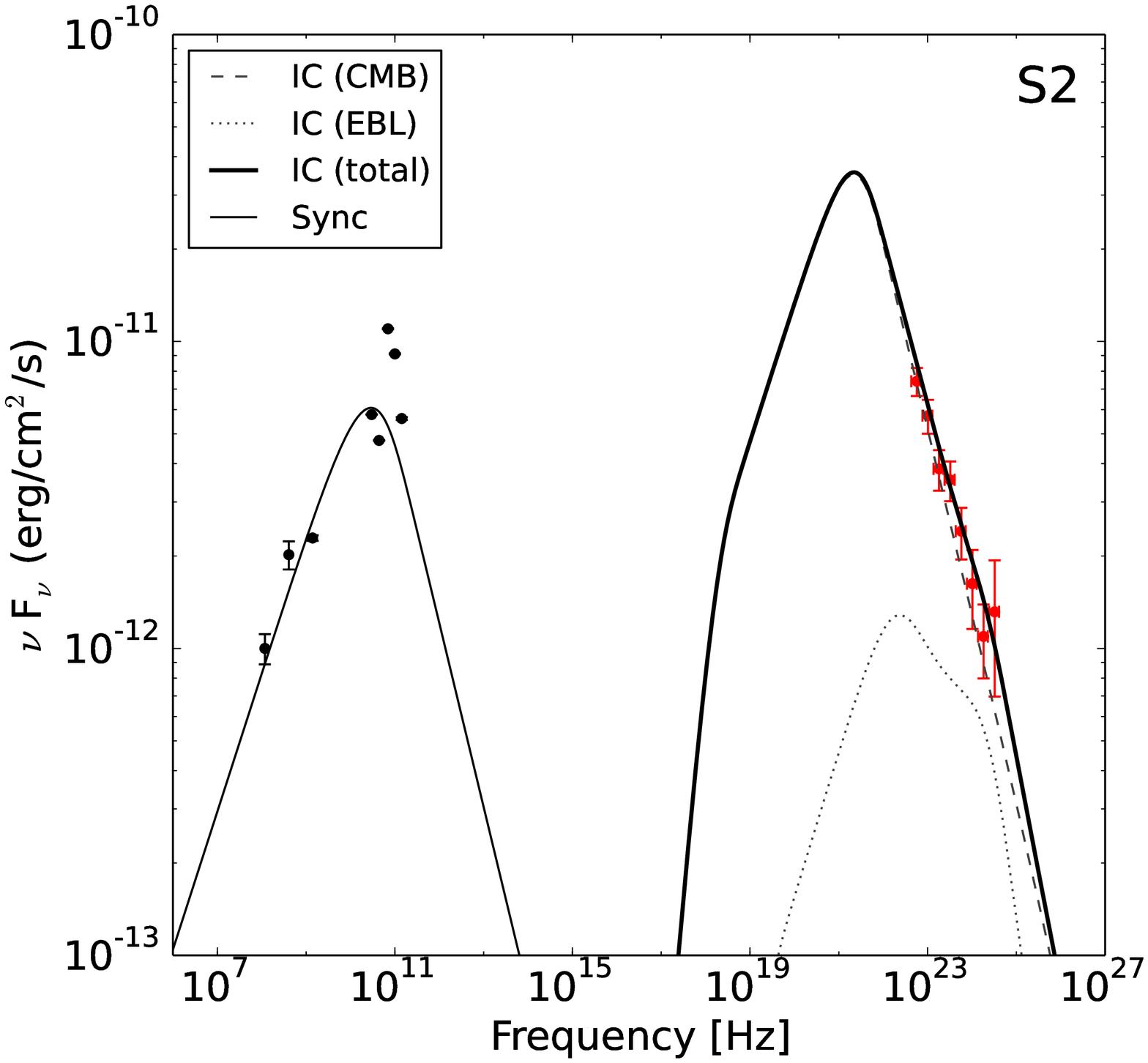}
\includegraphics[angle=0,scale=0.29]{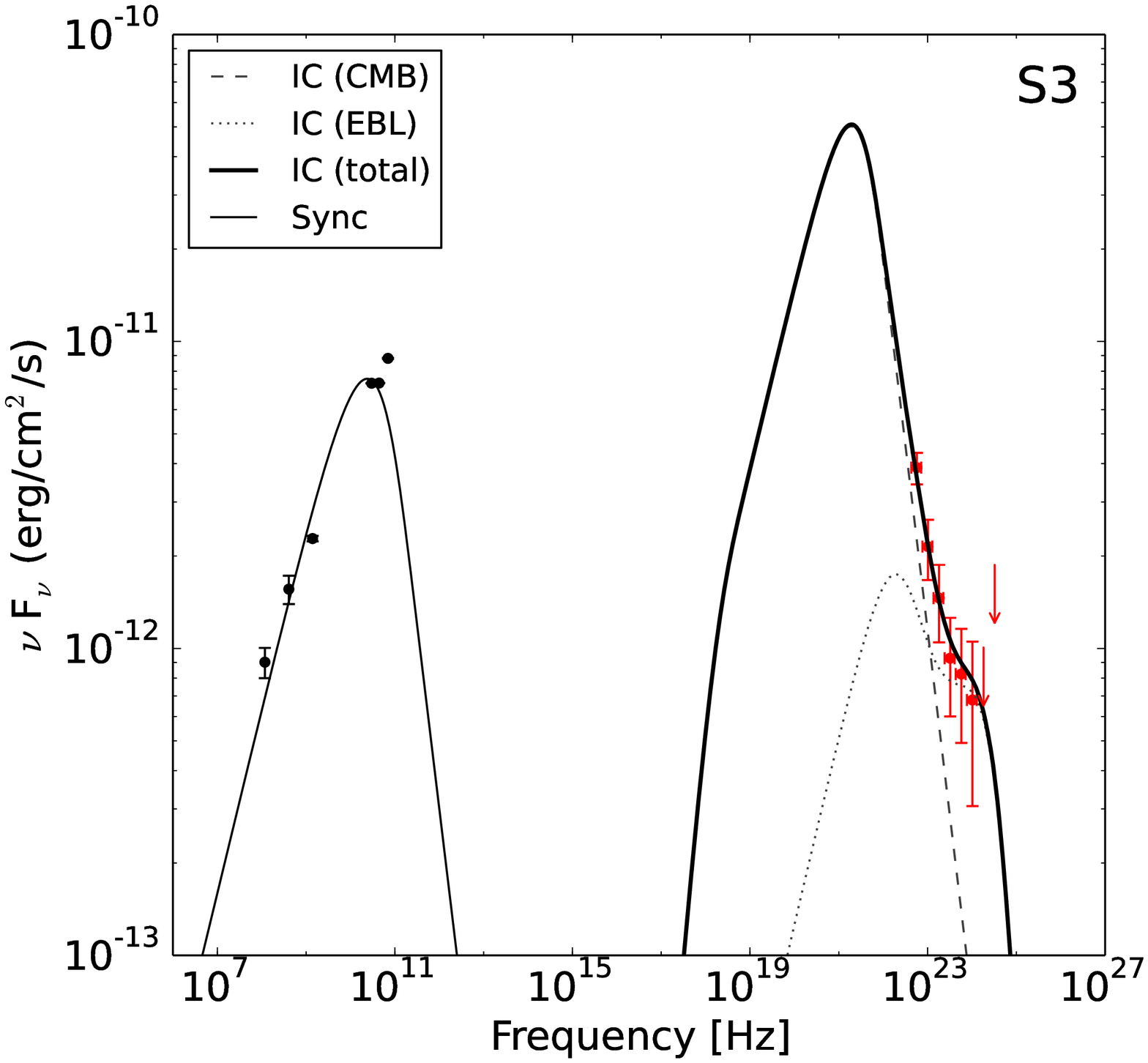}
\caption{Same as Fig.7 but for broken power-law electron distributions.}
\label{fig:SED_bpl}
\end{figure*}

\begin{figure*}
\centering
\includegraphics[angle=0,scale=0.29]{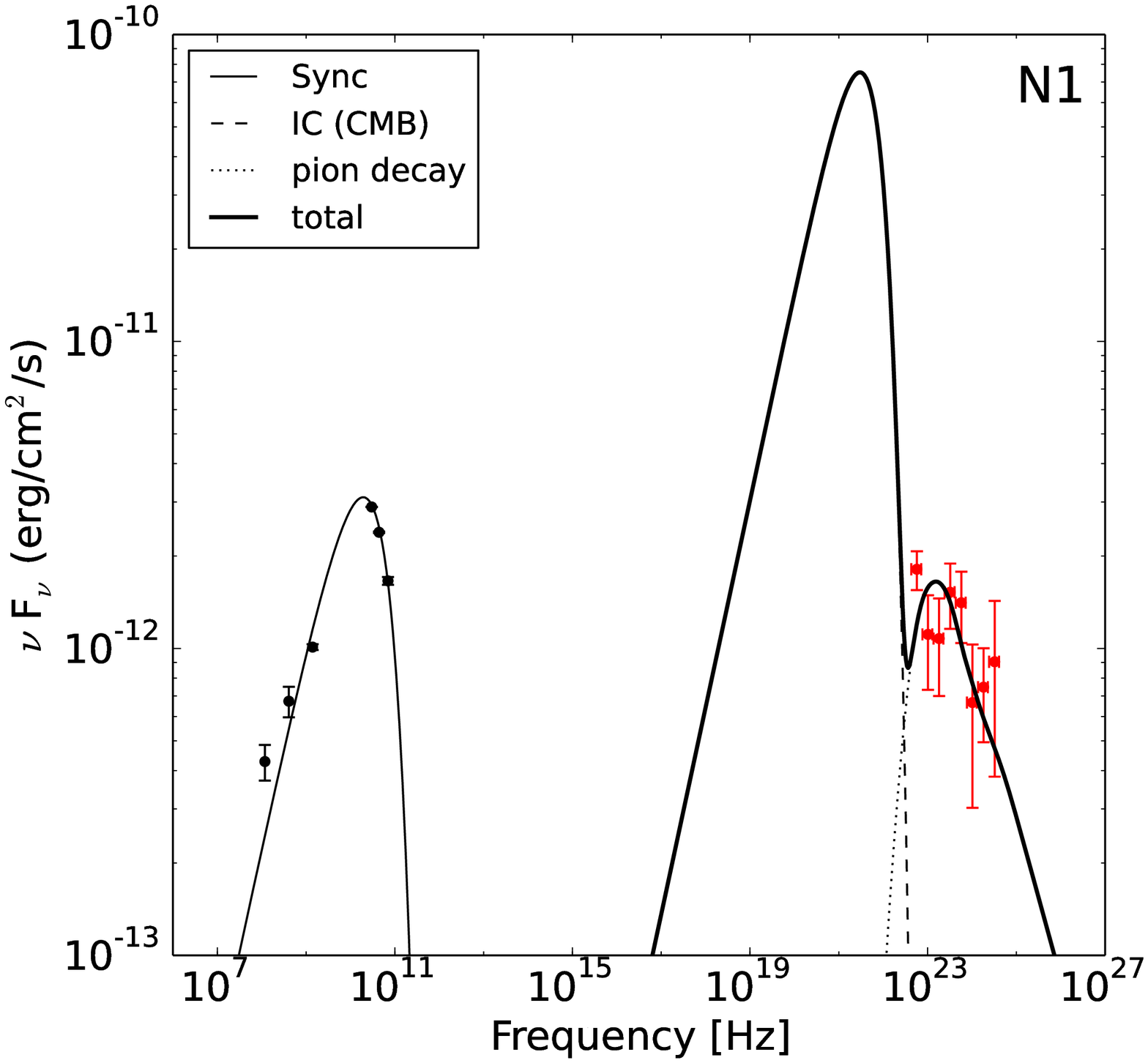}
\includegraphics[angle=0,scale=0.29]{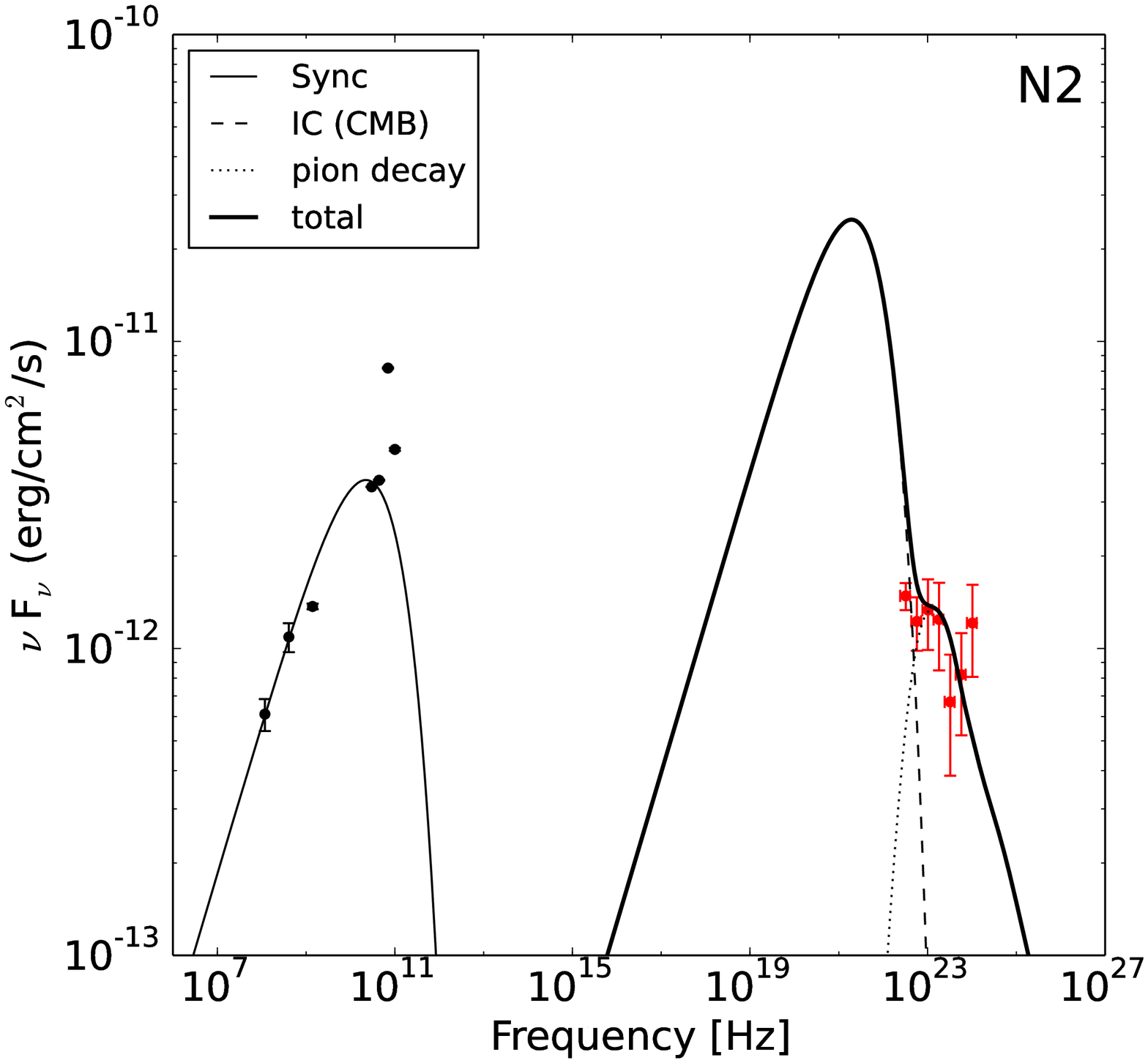}
\includegraphics[angle=0,scale=0.29]{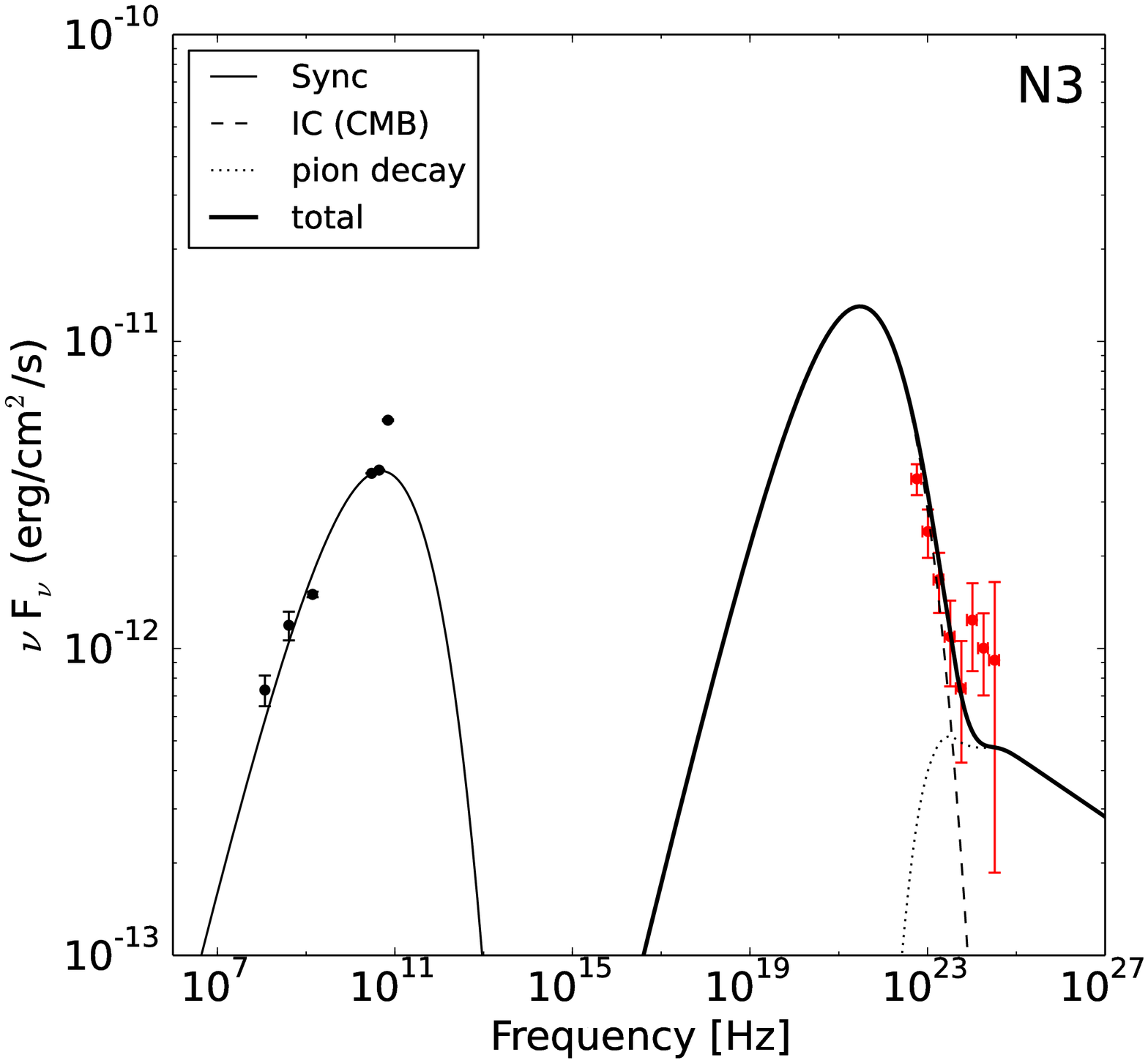}\\
\includegraphics[angle=0,scale=0.29]{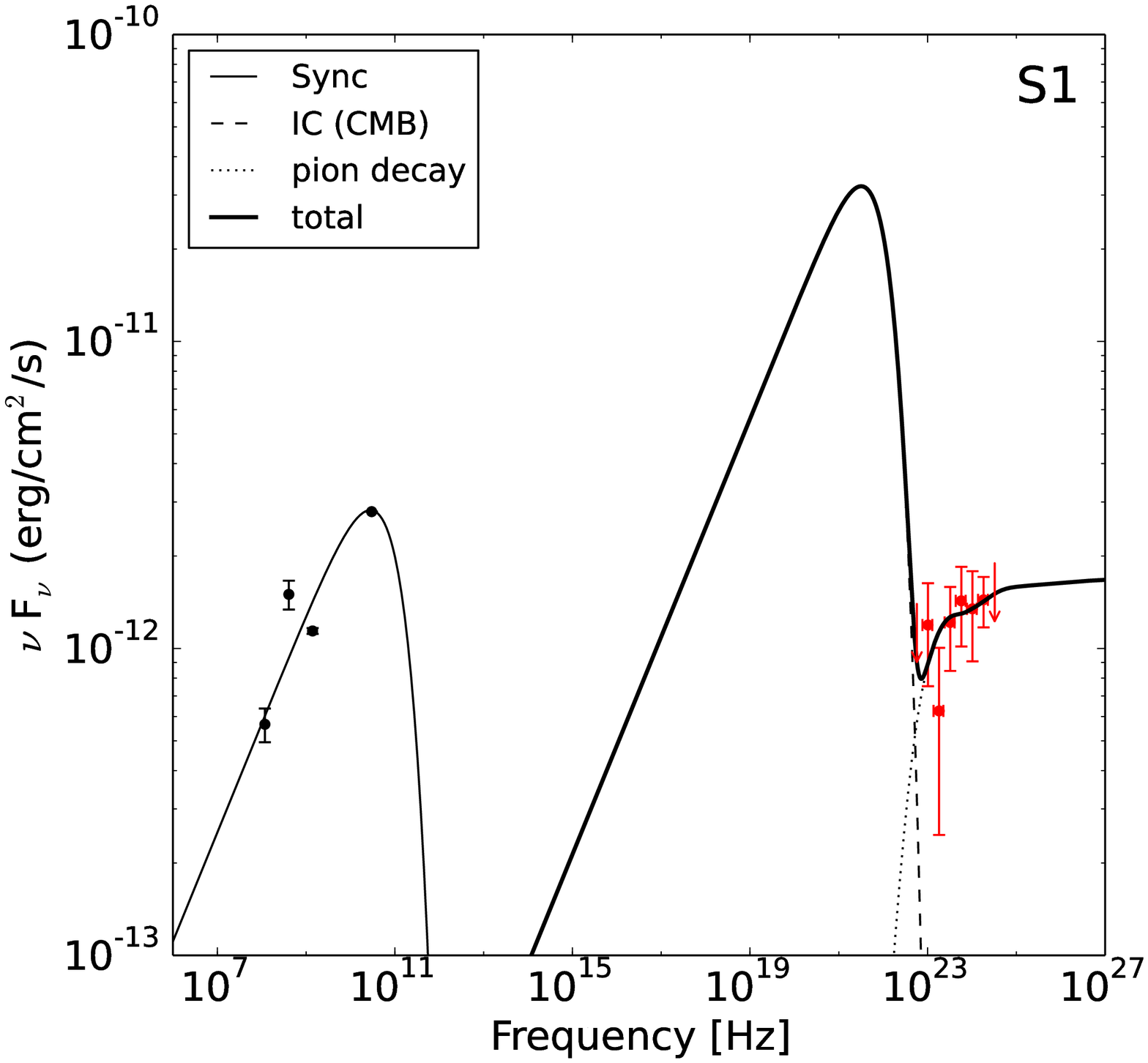}
\includegraphics[angle=0,scale=0.29]{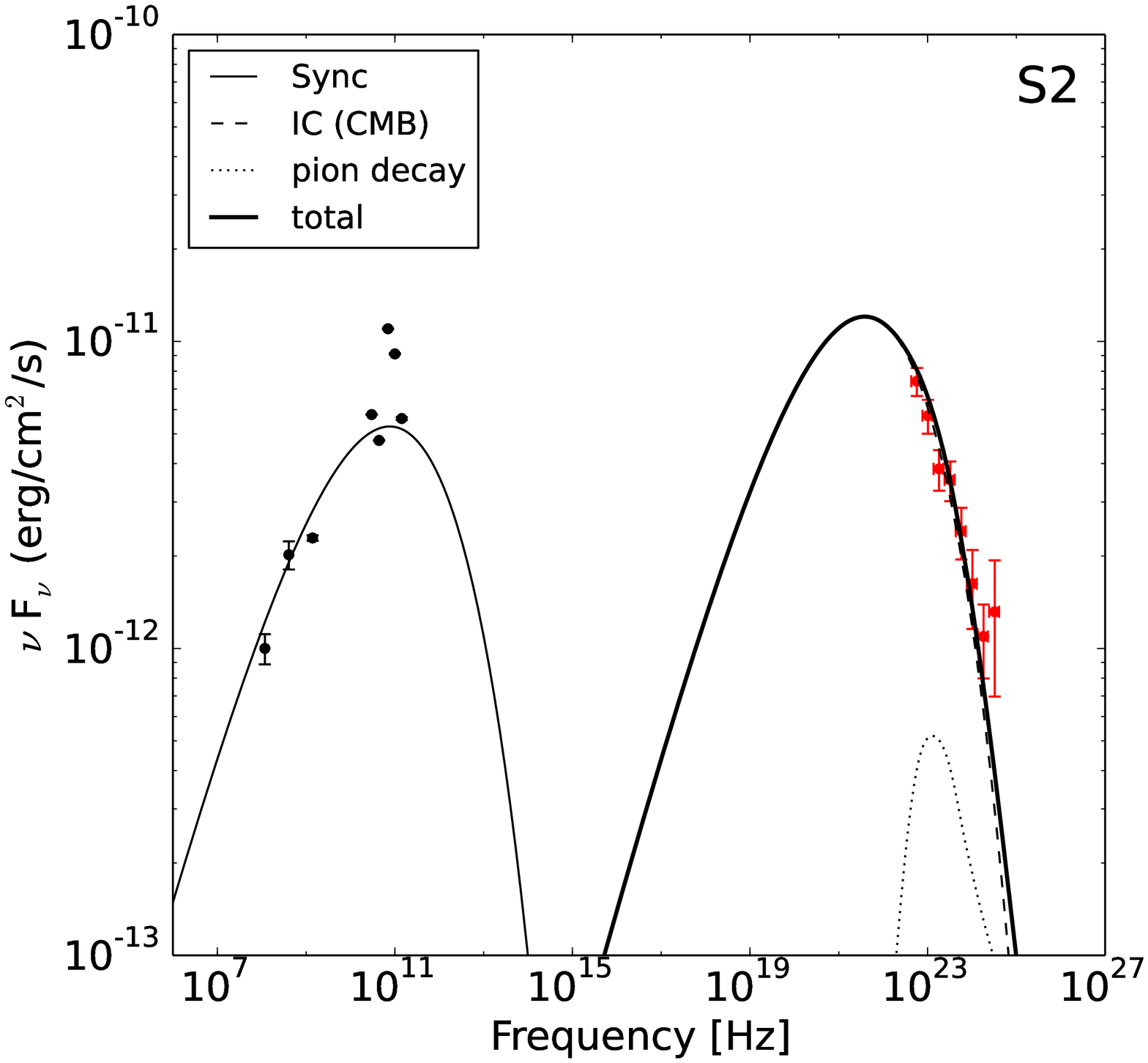}
\includegraphics[angle=0,scale=0.29]{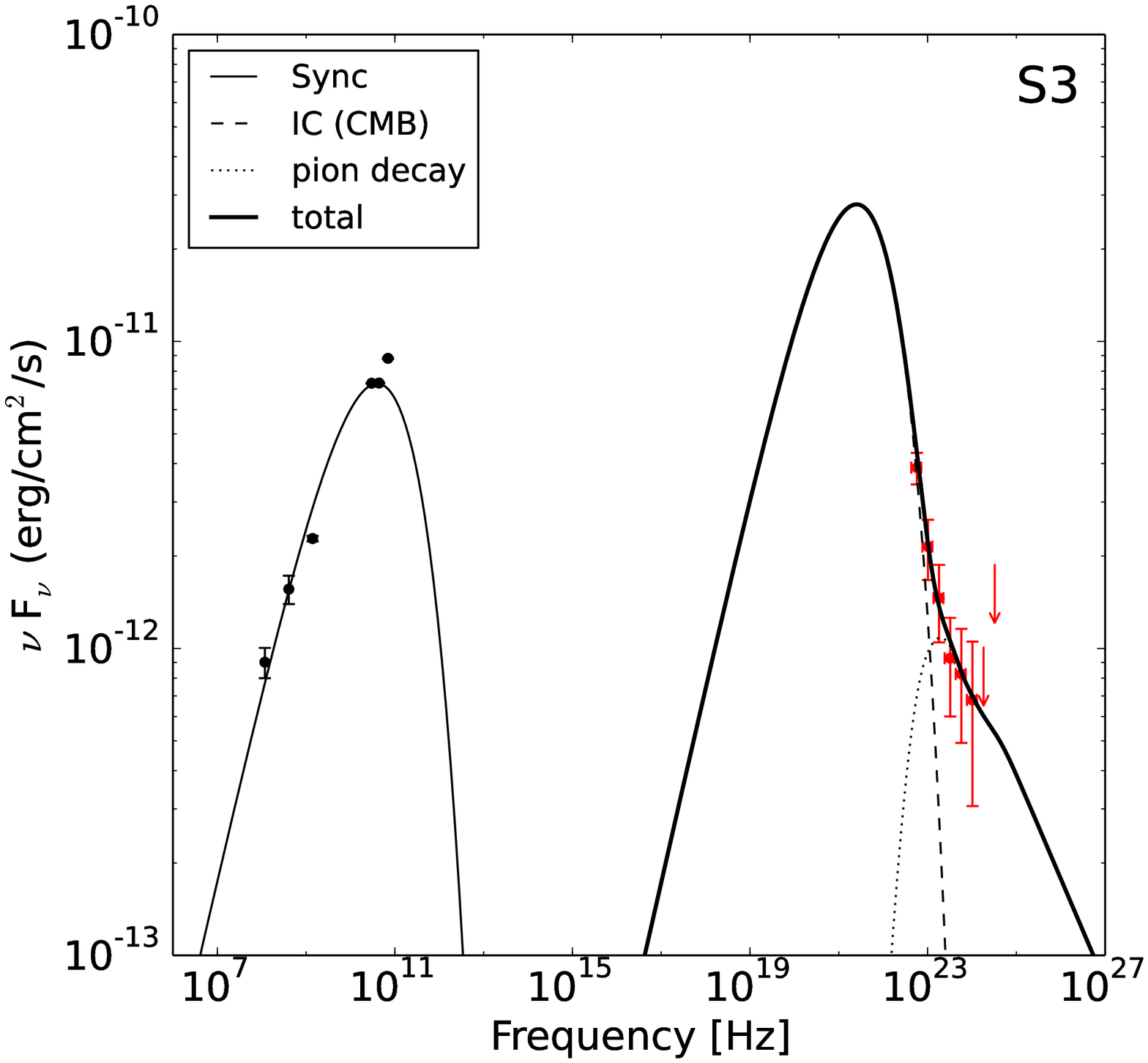}
\caption{Same as Fig.7 but for a hybrid model in which both leptonic and hadronic process contribute. }
\label{fig:SED_pp}
\end{figure*}

\begin{table*}
\centering
\caption{Summary of SED best-fit parameters in the leptonic-hadronic model.}
\label{table:6}
\begin{tabular}{ccccccc}
\hline\hline
Model components & N1 & N2 & N3 & S1 & S2 & S3 \\
\hline
$W_{\rm p}$~[$\times 10^{60}$ erg]&1.04$\pm$0.12&0.78$\pm$0.11&0.42$\pm$0.19&1.4$^{+0.7}_{-0.4}$&0.19$^{+0.14}_{-0.08}$&0.61$\pm$0.10\\
$\alpha$&2.59$\pm$0.12&2.7$\pm$0.2&2.16$^{+0.20}_{-0.11}$&2.05$\pm$0.12&2.7$^{+0.4}_{-0.3}$&2.48$^{+0.08}_{-0.06}$\\
\hline
\end{tabular}
\end{table*}

We should note that in this study we found,  in contrast to the statement of our previous paper \citep{2012A&A...542A..19Y},
that the EBL photons 
appear important (except for  the region S1), as  target photons for the IC scattering,  to fit the 
$\gamma$-ray data.  The reason is that  now the additional Planck data provide 
stringent constraints on the cutoff regions  of the electron spectrum.

For all  regions, except for S1,  the \gray spectra correspond to electrons from 
the post-cutoff region. Meanwhile the radio data are produced  by electrons from the pre-cutoff region. This follows from the essentially different indices of the radio and 
$\gamma$-ray spectra.  The region S1 is of special interest because of lack of any indication 
for a cut off in both the radio and \gray  spectra. In this region, the \gray and radio data points can be fitted with a pure power-law electron spectrum up to 1 TeV.  Another special feature of this region is that the  derived magnetic field is about $10~\rm \mu G$, which is much higher than  in other regions.
This value exceeds by an order of magnitude the strength of the magnetic field typically assumed for the radio lobes \citep[see e.g.][]{2015MNRAS.446.3478M}.


The dynamical ranges of both the radio and $\gamma$-ray data points are relatively small, 
therefore the power-law electron spectrum with a cutoff is not an unique explanation of the data. 
For example, the broken  power-law function, given in the form
\begin{equation}
\begin{split}
&N(E) = A~(\frac{E}{E_{0}})^{-\alpha_{\rm 1}} , ~E<E_{\rm break} \\
&N(E) = A~(\frac{E}{E_{0}})^{-\alpha_{\rm 2}} ~(\frac{E_0}{E_{\rm break}})^{\alpha_{\rm 1}-\alpha_{\rm 2}}, ~E<E_{\rm break},\\
\end{split}
\end{equation}
can fit the radio and $\gamma$-ray data equally well. 
The results are shown in Fig.\ref{fig:SED_bpl}. The best-fit parameters are summarised in Table \ref{table:5}.  We did not apply this electron distribution to the region S1 since the latter is explained by  a pure power-law spectrum. The differences in the indices before and after the break are significantly larger than 1. This implies that the break cannot be  a result of radiative 
cooling, but rather is a characteristic feature of the  acceleration spectrum.


As mentioned above, at low energies, $E \leq 1$~GeV,  the $\gamma$-ray data can be
explained only by  directly accelerated electrons. However, we cannot exclude a significant  
contribution by a hadronic component to the overall $\gamma$-ray emission. Moreover, 
an additional  hadronic component helps us to improve the fit of $\gamma$-ray spectra. In particular, the hadronic $\gamma$-ray emission  could be considered as an alternative to the IC scattering on the EBL photons.   Such an attempt to fit the radio and $\gamma$-ray  SEDs successfully, with an involvement of an additional hadronic component, is demonstrated in  Fig.\ref{fig:SED_pp}. 
In this case IC scattering  from CMB contributes to the low energy part of \grays, while the $\pi^0$ decays contribute to the high energy tail. This is similar, to some extent,  to the modelling of the radio lobes of Fornax A in \citet{2015MNRAS.446.3478M}, where the X-ray flux is due to the 
IC scattering, and  $\gamma$-rays are from the $\pi^0$-decays.

To reduce the number of free parameters, in the ``IC+$\pi^0$'' model we fix the magnetic field and electron spectrum to the best-fit values from the pure leptonic models described above.  The only exception 
was the peculiar S1 region, for which we fixed the magnetic field to the value of $1 \mu$G, i.e. by a factor of 10 smaller than in the pure IC scenarios.  We adopt the parametrisation of neutral pion decay described in \citet{kafexhiu14} in the $\pi^0$ model calculation. 
We also fix the  gas density, $n=10^{-4}~\rm cm^{-3}$.  Then,  the remaining two free parameters are the spectral index $\alpha$, and the total energy in protons, $W_{\rm p}$.   The  derived 
values of these parameters are presented in Table \ref{table:6}.  The power-law indices of the proton spectra in the different regions are similar with an average value close to 2.5. The only exception is the region S1, where the photon spectrum is very hard with $\alpha \sim 2$. The total energy in 
relativistic protons in  S1 is also different; it is  significantly higher than in other regions. On the other hand,   the additional hadronic component permits the reduction of the magnetic field to a nominal 
value of about $1 \mu$G.  Finally, in the south lobe we see a hardening of the proton spectrum. Interestingly, such an effect has also been seen in Fermi bubbles  \citep{su12, fermi_fb, yang14}, which are two giant, $\sim 10$~kpc scale, $\gamma$-ray structures belonging to our Galaxy.

\section{Conclusion and discussion}\label{sec:conc}
We have  analysed  the SEDs of the giant lobes of Cen A across a wide range of energies. The 
presented results increase  the significance of  the $\gamma$-ray detections reported before,  
and, more importantly, significantly extend the $\gamma$-ray spectrum down to 60~MeV and 
up to 30 GeV.  This allows  us to make rather  
robust conclusions regarding the origin of different components of  the $\gamma$-ray emission. 

We confirm the different morphologies of the giant lobes in $\gamma$-rays  and at radio frequencies. 
This  can be explained by the fact that the morphology of synchrotron radiation
is strongly affected by the spatial distribution of the magnetic field. Also, 
the electrons responsible for  \gray emission have higher energies than the electrons producing  synchrotron emission in the lobes.  To minimise the energy gap between the electrons responsible for the IC and synchrotron electrons,  we further analysed the high frequency \planck data.   

We divided both lobes into three regions and found significant spectral variations between regions. The power-law shape of the SED down to 100 MeV provides evidence against the hadronic origin of the emission.  On the other hand, the extension of $\gamma$-ray emission well beyond 10~GeV, and the inclusion of the \planck data permits more comprehensive spectral studies and broadband modelling of the SEDs.  All regions in the south and north lobes, except for  the region S1,
can be  naturally explained within a pure leptonic model in which the $\gamma$-rays are produced because of the IC scattering of electrons on the CMB photons with a non-negligible contribution from the EBL photons.  The  magnetic field and the total energy in relativistic electrons in the lobes, which are derived from a comparison of the SED modelling and the \fermi and \planck data, are about $1 ~\mu$G and  $W_{\rm e} \approx  6 \times 10^{57}$erg.  

The region S1 has very different radiation characteristics compared to the other regions. 
This is the only region where the radio and $\gamma$-ray components have the same spectral index, and a single power-law electron spectrum is required without 
a break or cutoff up to energies of 1~TeV.   
Although the SED of the region S1 can also be explained within a simple leptonic model, it however 
requires an unusually large magnetic field, $B \simeq 13 \mu$G, which is an order of magnitude larger than the average field in the lobes. On the other hand, the total energy of electrons in S1 is much smaller than in other regions, 
by a factor of 3 to 15. Thus, the ratio of pressures due to 
the magnetic field and relativistic electrons  in S1  differs  by 2 to 3 orders of magnitude from the 
average value in the lobes.  

An alternative explanation for the  peculiar features of S1 could be the effect of a 
non-negligible contribution of a new radiation component, presumably of hadronic origin. This contribution, on the top of the IC 
component, should become significant only at high energies, therefore does not  contradict the 
above claim that $\gamma$-rays  below 1~GeV should be dominated by the IC scattering  
of electrons.

 Indeed if we ignore the IC component from EBL, which is poorly constrained, 
 the \gray spectra in all regions can be interpreted as a combination of two 
 components: IC scattering on CMB photons and hadronic $\gamma$-rays from the pion decays. The derived total proton energy budget  is of the order of $5\times 10^{60}\rm erg$, which is consistent with the estimation in  \citet{2012A&A...542A..19Y}. Such energy could only be accumulated on a  timescale as long as $10^9$ yrs, assuming  an  injection rate of the order of $10^{44}~\rm erg$.  The diffusion coefficient of cosmic rays in this case can be estimated from the condition of their propagation to distances of order of 100~kpc: 
$D \sim R^2/t \sim 3\times10^{30}\frac{R}{100~\rm kpc} ~\rm cm^2/s $.  The hardening 
 of the spectrum of cosmic rays in the south lobe  is  similar to the spectral hardening towards 
 the edge of Fermi bubbles \citep{su12, fermi_fb, yang14}, which may be related to the energy dependent propagation of cosmic rays.   
 
A possible problem of the leptonic-hadronic model applied to the lobes of Cen A is the 
 huge overall energy required in relativistic protons. It exceeds the
 total energy in the magnetic field by two orders of magnitude. This problem, however, can be reduced if we assume that 
 the $\gamma$-ray production at p-p collisions  takes  place primarily in the filamentary structures of the lobes. This is similar to the idea in \citet{crocker14},  who proposed that the collapse of thermally unstable plasma inside Fermi bubbles  can lead to an accumulation of cosmic rays and magnetic field into localised filamentary condensations of higher density gas. If this is the case  in the giant radio lobes of Cen A as well,  the required  energy budget in CRs can significantly reduce the required  cosmic ray energy budget. 

Concerning the relativistic electrons, their  origin remains a  mystery. The huge size of the giant lobes makes it impossible to transport the relativistic electrons from the core of Cen A. More specifically, the SED fitting results show that we need uncooled electrons of energy up to 50 GeV in the north lobe. The cooling time of 50 GeV electrons can be estimated as $t_{\rm cool} \sim\frac{ 2\times10^{19}}{w \gamma} ~\rm s~\sim25 ~\rm Myr$, where $w$ is the energy density of the ambient radiation and magnetic fields in unit of $\rm eV~cm^{-3}$ and $\gamma$ is the Lorentz factor of the electrons. The propagation length of electrons during  the cooling time is then $l \sim 30~\rm kpc (\frac{D}{10^{29}~\rm cm^2/s})^{0.5}$, i.e.  far less than the size of the lobe. The only solution is the {\it in situ} acceleration of the electrons, such as the stochastic acceleration in a turbulent magnetic field. 
\section*{Acknowledgements}

\bibliographystyle{aa}
\bibliography{CenA}

\end{document}